\documentclass[prb,twocolumn,floatfix,footinbib,bibnotes,longbibliography,superscriptaddress]{revtex4-2}
\usepackage{physics}
\usepackage{graphicx}
\usepackage{float}
\usepackage{amsmath}
\usepackage{amssymb}
\usepackage{hyperref}
\hypersetup{colorlinks=true,citecolor=blue,linkcolor=blue,urlcolor=blue}
\usepackage{xcolor}
\usepackage[normalem]{ulem}

\newcommand{\ci}{\mathrm{i}}

\newcommand{\redtext}[1]{\textcolor{black}{#1}}

\begin{document}
\title{Universal non-equilibrium dynamics of pure states and density-dependent thermalization in Sachdev-Ye-Kitaev model}

\author{Rishik Perugu}
\email{rperugu@uci.edu}
\affiliation{Department of Physics and Astronomy, University of California, Irvine, California 92697, USA}
\affiliation{Undergraduate Programme, Indian Institute of Science, Bangalore 560012, India}
\author{Arijit Haldar}
\email{arijit.haldar@bose.res.in}
\affiliation{S. N. Bose National Centre for Basic Sciences, JD Block, Sector-III, Salt Lake City, Kolkata - 700 106, India}
\author{Sumilan Banerjee}
\email{sumilan@iisc.ac.in}
\affiliation{Centre for Condensed Matter Theory, Department of Physics, Indian Institute of Science, Bangalore 560012, India}

\date{\today}

\begin{abstract}
Non-equilibrium dynamics of unentangled and entangled pure states in interacting quantum systems is crucial for harnessing quantum information and to understand quantum thermalization. We develop a general Schwinger-Keldysh (SK) field theory for non-equilibrium dynamics of pure states of fermions. We apply our formalism to study the time evolution of initial density inhomogeneity and multi-point correlations of pure states in the complex Sachdev-Ye-Kitaev (SYK) models. We demonstrate a remarkable universality in the dynamics of pure states in the SYK model. We show that dynamics of almost all pure states in a fixed particle number sector is solely determined by a set of universal large-$N$ Kadanoff-Baym equations. Moreover, irrespective of the initial state  the site- and disorder-averaged Green's function thermalizes instantaneously, whereas local and non-local Green's functions have finite thermalization rate. \redtext{We provide understanding of our numerical and analytical large-$N$ results through random-matrix theory (RMT) analysis}.
Furthermore, we show that the thermalization of an initial pure product state in the non-interacting SYK$_2$ model is independent of fermion filling and an initial density inhomogeneity decays with weak but long lived oscillations due to dephasing. In contrast, the interacting SYK$_{q\geq 4}$ model thermalizes slower than the non-interacting model and exhibits filling-dependent monotonic relaxation of initial inhomogeneity. For evolution of entangled pure states, we show that the initial entanglement is encoded in the non-local and/or multi-point quantum correlations that relax as the system thermalizes.
\end{abstract}

\maketitle

\section{Introduction}
The study of non-equilibrium dynamics in quantum many-body systems is a topic of great current interest. In particular, the time evolution of various kinds of pure states, leading to generation of entanglement and ensuing quantum correlations, has enormous relevance for understanding thermalization and quantum dynamics in condensed matter systems \cite{NandkishoreAnnRev2015, Altman2015, PolkovnikovRMP2011}, black holes in quantum gravity \cite{Sekino2008,MSS2016,Haehl2017,Almheiri2021}, and control and manipulation of quantum states for quantum computing \cite{Divincenzo1995,Riera2012}. However, theoretical description of the far-from-equilibrium time evolution in interacting quantum many-body systems, especially starting from a generic initial pure state, is extremely challenging. Non-equilibrium Schwinger-Keldysh (SK) field theory \cite{Kamenev2011,Stefanucci2013} and its application to interacting systems under various approximations \cite{Kamenev2011,Stefanucci2013,Berges2003,Berges2004,Aoki2014}, generally considers time evolution of initial mixed state. Additionally, these approaches often assume adiabatic evolution from equilibrium states of a non-interacting system.  As a result, studies of the dynamics of pure states are often limited to brute-force numerical computations, which are typically restricted to small systems, or states with low entanglement, described, e.g., by matrix product state or related ansatz \cite{Vidal2004, White2004, SCHOLLWOCK2011, PAECKEL2019}. Consequently, theoretical works on the exact time evolution of generic unentangled and entangled pure states for strongly interacting systems in the thermodynamic limit are extremely rare.

In recent years, certain class of models of strongly interacting fermions, namely the Sachdev-Ye-Kitaev (SYK) and related models \cite{SY1992, kitaevtalk2015, SachdevSYK2015, SYK2016,ChowdhurySYKReview,Gu2017,BAmodel2017,Jian2017,Song2017,Davison2017,Zhang2017,Chowdhury2018,Haldar2018,Haldar2018a,Jian2018,Esterlis2019,Kim2021,Patel2023}, have provided a paradigm to study non-trivial thermodynamics, equilibrium and non-equilibrium dynamics exactly in the large-$N$ or the thermodynamic limit. The SYK model is maximally chaotic at low temperature, i.e., it saturates the Maldacena-Shenker-Stanford (MSS) bound \cite{MSS2016} on quantum chaos as the temperature $T\to 0$. The ground state of the SYK model is a non-Fermi liquid (NFL) lacking Landau's quasiparticle description \cite{ChowdhurySYKReview}. The model exhibits a non-zero residual entropy at zero temperature in the large-$N$ limit \cite{kitaevtalk2015, SachdevSYK2015, SYK2016}. These, and other associated features \cite{kitaevtalk2015,SYK2016,Kitaev2018}, suggest intriguing connections of SYK model to strange metals in condensed matter systems and black holes in quantum gravity. However, despite the analytical tractability, studies of non-equilibrium dynamics even in the large-$N$ limit of SYK models have been limited to evolution of mixed states, e.g., after a quantum quench \cite{Eberlein2017, Arijit2020, BhattacharyaSYKQuench, jaramillo2024thermalizationclosedsachdevyekitaevthermodynamic,Almheiri2019,Samui2021,Larzul2022,Louw_manySYK,Louw2023,Hosseinabadi2023}. Broadly, these studies have explored the distinctions between Fermi and non-Fermi liquids in terms of their thermalization dynamics. In some cases, non-equilibrium dynamics, such as the nonlocal correlations and the growth of entanglement \cite{kourkoulou2017purestatessykmodel,ZhangPureStateSYKdynamics, Sohal2022}, of certain specially constructed pure states of Majorana fermions, namely the Kourkoulou-Maldacena (KM) states, has been studied in the Majorana version of the SYK model. Also, there have been a few numerical studies \cite{Bandyopadhyay2023,Dieplinger2023} of non-equilibrium evolution and thermalization of both mixed and pure states in SYK models for finite $N$.

Here we develop a general Schwinger-Keldysh non-equilibrium field theory method for the time evolution of arbitrary pure states of fermions. We apply the formalism to study the thermalization of unentangled and entangled pure states exactly in the large-$N$ limit of the SYK model constituting complex fermions \cite{SachdevSYK2015,Davison2017}. Unlike the \emph{Majorana SYK model}, the particle-number conservation in the \emph{complex SYK model} allows us to study the thermalization dynamics for high-energy pure states as a function of fermion density or filling. Moreover, our method enables us to study the non-trivial time-evolution of non-local two and multi-point quantum correlations as imprints of entanglement of initial states during the approach to thermalization. We find that the time evolution of all generic pure states, not correlated with SYK Hamiltonian, for a fixed particle number sector is entirely obtainable from a single Green's function and associated universal large-$N$ saddle point equations. This remarkable universality among far-from equilibrium dynamics of distinct pure states emerges due to lack of correlation between any given initial pure state and random couplings of the SYK Hamiltonian. Below, we provide an overview of our SK field theory formalism for pure states and the results obtained from the application of the formalism to SYK model.

\subsection{Overview of the results} \label{sec:Overview}
\begin{figure}[ht]
    \centering
    \includegraphics[width=\linewidth]{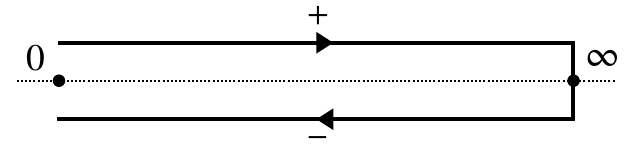}
    \caption{Schwinger-Keldysh contour $\mathcal{C} = [0, \infty) \cup (\infty, 0]$, for an initial pure state, consisting of a forward $(+)$ and a backward $(-)$ branch, between $0$ and $\infty$.}
    \label{fig:KeldyshContour}
\end{figure}

To construct the SK non-equilibrium field theory for evolution of pure state, we first consider initial pure product states  $|\Psi(0)\rangle=|n\rangle \equiv |n_1,n_2,\cdots,n_N\rangle$ where $n_i=0,1$ is the occupation of suitable single-particle basis, indexed by $i=1,2,\cdots,N$, denoting, for example, fermion sites or flavors in case of SYK model. The product occupation state is far away from any equilibrium state for time evolution under a generic Hamiltonian. The initial density matrix $\rho(0)=|n\rangle\langle n|$ is then evolved with many-body Hamiltonian $\rho(t)=e^{-\mathrm{i}\mathcal{H}t}\rho(0)e^{\mathrm{i}\mathcal{H}t}$ ($\hbar=1$), and the evolution is described via a SK generating function or coherent-state path integral $Z=\mathrm{Tr}[\rho(t)]=\int \mathcal{D}(\bar{c},c)e^{\mathrm{i}S}\langle c(0+)|\rho(0)|-c(0-)\rangle$ over the fermionic Grassmann fields $\{\bar{c}_i(t\pm),c_i(t\pm)\}$ ($0\leq t<\infty$) on the SK contour with forward ($+$) and backward ($-$) branches [Fig.\ref{fig:KeldyshContour}]. Crucially, information about the initial state enters in the SK path integral through the coherent-state matrix element $\langle c(0+)|\rho(0)|-c(0-)\rangle$. The general strategy to deal with initial condition in SK field theory, as employed for the usual initial thermal density matrices \cite{Kamenev2011,Stefanucci2013} or other non-thermal initial states \cite{Berges2003,Berges2004}, is to exponentiate the matrix element and include it in a modified action $S\to S+S_{in}$. In the same vein, we write the matrix element $\langle c(0+)|n\rangle\langle n|c(0-)\rangle$ as an exponential by introducing auxiliary static Grassmann source fields through an elementary, albeit crucial, integral identity. The auxiliary Grassmann fields couple linearly to the dynamical fermion fields $(\bar{c},c)$ at time $t=0$. The auxiliary Grassmann fields can be integrated out at a suitable intermediate stage or at the end of the calculation depending on the model and approximation used.

The method is easily generalized further for an arbitrary and, in general, entangled initial pure state $|\Psi(0)\rangle=\sum_n C_n|n\rangle$, where all the occupation states $|n\rangle$ in the superposition have the same total number of fermions.
Our method, different from several earlier works \cite{Berges2003,Berges2004} with non-thermal initial conditions in SK field theory, is similar in spirit to that in refs.\cite{Chakraborty2019,Chakraborty2020}. In the latter, bosonic source fields, coupled to the bilinear of the dynamical fermion field, are used to exponentiate the initial density matrix element. As a result, all the calculations at the intermediate steps of the SK field theory are performed in the presence of the source fields, and eventually an extensive number of derivatives with respect to the source fields need to be performed at the final step of the calculations. As we demonstrate for the case of SYK model, our method circumvents this step of differentiations, which is technically challenging, in general, for interacting systems.

\begin{figure}
    \centering
    \includegraphics[width=0.48\textwidth]{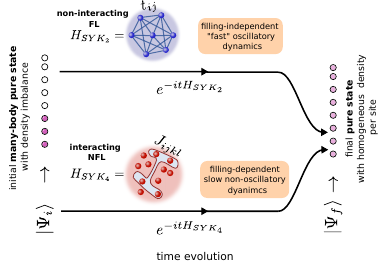}
    \caption{{\bf Far-from equilibrium dynamics of pure states and thermalization in SYK models:} Time evolutions of an initial pure state $|\Psi_i\rangle$ with density inhomogeneity towards a homogeneous pure state $|\Psi_f\rangle$ under unitary dynamics in non-interacting SYK$_2$ and interacting SYK$_4$ models show contrasting behavior. The relaxation of inhomogeneity in the SYK$_2$ model is insensitive to fermion filling and oscillatory, whereas the SYK$_4$ model exhibits slower filling-dependent thermalization without oscillations.}
    \label{fig:schematic}
\end{figure}

We first apply our SK field theory for pure states to study the time evolution of occupation states $|n\rangle$ with fermion filling or density $f=(1/N)\sum_in_i$ in the SYK$_q$ model with random $(q/2)$-body ($q$ even) interactions. In particular, we focus on interacting $\mathrm{SYK}_4$, and contrast its thermalization dynamics with that of non-interacting $\mathrm{SYK}_2$ model (see Fig.\ref{fig:schematic}). The latter describes fermions with all-to-all random hoppings on $N$ sites. To this end, exact large-$N$ saddle point equations for disordered averaged fermionic Green's functions, where the memory of the initial state only appear through filling, are derived from our SK field theory. The self-consistent saddle-point equations are written as a set of causal Kadanoff-Baym (KB) equations and are numerically integrated to obtain the non-equilibrium evolution of the fermionic Green's functions. \redtext{We also obtain exact analytical solutions of the KB equations for the $\mathrm{SYK}_2$ and large-$q$ $\mathrm{SYK}_q$ models, as well as asymptotic late-time solution for general $q\geq 4$. These numerical and analytical solutions of the KB equations enable us to} obtain the evolution of the Green's functions $G_e(ts,t's')$ ($s,s'=\pm$) for the \emph{initially empty sites} on the SK contour. We show that $G_e(ts,t's')$ determines the Green's function, $G_f(ts,t's')$ of the \emph{initially filled sites}, and the collective large-$N$ Green's function $G(ts,t's')=(1-f)G_e(ts,t's')+fG_f(ts,t's')$. The latter is obtained by averaging local Green's function over all the sites. We find that the large-$N$ Green's function thermalizes instantaneously, i.e., $G(ts,t's')=G^{ss'}(t-t')$ for any $t,t'$, even though the local Green's functions exhibit non-trivial relaxation with finite thermalization rate. The instantaneous thermalization of $G$ arises since almost any pure state is uncorrelated with the random couplings of the SYK model, as we verify through a random matrix theory (RMT) analysis. \redtext{The RMT analysis establishes a connection between site and disorder-averaged local Green's function and a quantity similar to spectral form factor (SFF) \cite{Brezin1997} for the dynamics of a generic pure state under random-Hamiltonian evolution.}

From the perspective of eigenstate thermalization hypothesis (ETH) \cite{Deutsch1991,Srednicki1994,Srednicki1999}, the product states $|n\rangle$, having energy (expectation) at the middle of the many-body energy spectrum of the SYK model, corresponds to \emph{infinite-temperature states}. Thus, we probe thermalization of very high-energy pure states, in contrast to the low-energy Fermi and non-Fermi liquid regime studied in earlier works on non-equilibrium dynamics of mixed states in SYK models \cite{Eberlein2017, Arijit2020, BhattacharyaSYKQuench, jaramillo2024thermalizationclosedsachdevyekitaevthermodynamic,Almheiri2019,Samui2021,Larzul2022,Louw_manySYK,Louw2023,Hosseinabadi2023}. Remarkably, even for such high-energy states $|n\rangle$, we find striking differences between the large-$N$ thermalization dynamics of the non-interacting $\mathrm{SYK}_2$ and interacting $\mathrm{SYK}_{q\geq 4}$ models. For $\mathrm{SYK}_2$, the initial density inhomogeneity or \emph{imbalance} between filled and unfilled sites in $|n\rangle $ exhibits oscillatory relaxation independent of density $f$. In contrast, $\mathrm{SYK}_{q\geq 4}$ displays non-oscillatory and strongly density-dependent relaxation of imbalance. We benchmark our large-$N$ results for the imbalance dynamics in the $\mathrm{SYK}_2$ model with the disorder-averaged imbalance obtained via exact diagonalization of the random hopping matrix over many realizations for finite but large $N$. \redtext{More notably, using single-particle RMT, we derive an analytical expression for the imbalance dynamics that matches exactly with the analytical solution of the large-$N$ KB equations in the $\mathrm{SYK}_2$ model.} 

\redtext{Furthermore, the exact solution of the large-$N$ KB equations for $\mathrm{SYK}_q$ model in the large-$q$ limit and asymptotic late-time solution for general $\mathrm{SYK}_{q\geq 4}$ enable} us to estimate a filling-dependent thermalization rate $\sim [f(1-f)]^{q/4-1/2}$ for the imbalance. As a result, the non-equilibrium Green's function of the empty (filled) sites $G_e$ ($G_f$) has a finite relaxation time even in the large-$q$ limit, even though the large-$N$ Green's function $G$ relaxes instantaneously for any $q$. Thus, the initial density inhomogeneity in our case allows us to probe thermalization in quantities, like the imbalance and $G_e$ ($G_f$), that are not accessible starting from homogeneous thermal mixed states, as in the earlier works \cite{Eberlein2017, Arijit2020, BhattacharyaSYKQuench, jaramillo2024thermalizationclosedsachdevyekitaevthermodynamic,Almheiri2019,Samui2021,Larzul2022,Louw_manySYK,Louw2023,Hosseinabadi2023}. \redtext{We show that analytically estimated thermalization rate exactly describes our numerical results for the relaxation of imbalance. Apart from the contrast in the filling dependence of the thermalization dynamics between the non-interacting $\mathrm{SYK}_2$ and interacting $\mathrm{SYK}_{q\geq 4}$ models, another major difference between the non-interacting and interacting cases appear in the late-time relaxation of the imbalance. The imbalance decays with a power law, along with the oscillations, in the $\mathrm{SYK}_2$ model. In contrast, the imbalance decays exponentially in the $\mathrm{SYK}_{q\geq 4}$ models. Given the contrasting thermalization dynamics} between $\mathrm{SYK}_2$ and $\mathrm{SYK}_4$ models, we study the crossover between the two limits in a model with both $\mathrm{SYK}_2$ and $\mathrm{SYK}_4$ terms. Even in this case, using short-time expansion of the Kadanoff-Baym equations, we are able to obtain estimate of the thermalization rate, that describes very well the crossover between the two limits observed in the numerical results.

Finally, we apply our SK field theory for an arbitrary initial state $|\Psi(0)\rangle=\sum_n C_n|n\rangle $ in a fixed particle number sector to describe the time evolution of initially entangled states in the SYK model. We show that, for any entangled initial state, the non-equilibrium evolution is obtainable in terms the Green's function $G_e(ts,t's')$ and the corresponding large-$N$ saddle-point equations, identical to that for pure product state $|n\rangle$. This universality among dynamics of pure states, entangled or unentangled, appears due to lack of correlation between any given $|\Psi(0)\rangle$ and random interactions in the SYK Hamiltonian. As a corollary, the site-averaged large-$N$ Green's function $G$ for any pure state also thermalizes instantly. Moreover, through a Kubo-Martin-Schwinger (KMS) condition or fluctuation dissipation ratio (FDR), we see that $G$ always corresponds to infinite temperature, irrespective of $|\Psi(0)\rangle$. Nevertheless, the entangled nature of the initial state leads to non-trivial time dependence of various correlations that are non-local in fermionic sites $i$. These quantum correlations, though all determined by various combinations of the same two-point function $G_e$, encode the information about non-trivial entanglement in the initial state and decay during the approach to thermalization. After the system thermalizes only the site-local correlations remains non-zero, while all the non-local correlations vanishes, as expected in the large-$N$ limit.



The paper is organized as follows. In Sec.\ref{models}, we describe the Hamiltonians for various SYK models that we use to study evolution of pure many-body states. In Sec.\ref{SKPureStateFormalism}, we develop our formalism utilizing the Schwinger-Keldysh method for studying the time evolution of a general pure state of fermions. We apply this formalism to the SYK models of Sec.\ref{models} and describe their large-$N$ SK field theories in Sec.\ref{SYKlargeN} for unentangled and entangled pure states. In Sec.\ref{Results}, we present the main results, \redtext{from both large-$N$ SK theory and RMT analysis}, for the time evolution of pure Fock state in the non-interacting, interacting, and mixed (non-interacting + interacting) SYK models. The time evolution of an initially entangled Bell pair state is also discussed here. Finally, we conclude our results and outline future directions in Sec.\ref{Conclusion}.
Relevant details such as derivation of KB equations, exact diagonalization of non-interacting models and large-$q$ analysis are provided in the appendices.

\section{Models}\label{models}

The Hamiltonian of the interacting SYK$_4$ \cite{SY1992, kitaevtalk2015, SachdevSYK2015, SYK2016} model is given by
\begin{equation}
    \mathcal{H}_4 = \frac{1}{(2N)^{3/2}} \sum_{ijkl} J_{ijkl} c^\dagger_i c^\dagger_j c_k c_l \label{eq:SYK4}
\end{equation}
where $c_i$’s are spinless complex fermions on sites $i=1,2, \dots, N$, and the two-particle interactions $J_{ijkl}$ are independent Gaussian random numbers with the mean zero and the variance $\expval{|J_{ijkl}|^2} = J_4^2$, while 
satisfying $J_{ijkl} = -J_{jikl} = -J_{ijlk} = J^*_{klij} $. The ground state of this model is a non-Fermi liquid, lacking  quasiparticle excitations \cite{ChowdhurySYKReview}. This model saturates the MSS bound for the Lyapunov exponent \cite{MSS2016} at low temperature, and thus is the \emph{fastest scrambler}. The model is also a fast thermalizer, with a thermalization rate $\propto T$, after a quench from a thermal mixed state \cite{Eberlein2017}. 

The non-interacting SYK$_2$ model is an all-to-all random hopping model whose Hamiltonian is given by
\begin{equation}
    \mathcal{H}_2 = \frac{1}{\sqrt{N}}\sum_{ij} J_{ij} c^{\dagger}_i c_j \label{eq:SYK2} 
\end{equation}
where the hopping matrix elements $J_{ij}$ are independent Gaussian random numbers with the mean $\expval{J_{ij}} = 0 $, and the variance $\expval{|J_{ij}|^2} = J_2^2$, while satisfying $J_{ij} = J_{ji}^*$. The model describes quantum mechanics of a $N\times N$ random matrix with single-particle chaos, e.g., in terms of single-particle level spacing. However, the model is non-chaotic from the many-body perspective, e.g., it has a Poisson statistics for many-body energy level spacings and zero Lyapunov exponent.

We also consider a model with both SYK$_2$ and SYK$_4$ terms, often dubbed the \emph{mass-deformed} SYK model in the literature \cite{ DelayedThermMassDefSYK, KComplexityChaosMassDefSYK, ThouslessTimeMassDefSYK}. The Hamiltonian of this model is given by
\begin{equation}
    \mathcal{H}_{24} =\frac{1}{(2N)^{3/2}} \sum_{ijkl} J_{ijkl} c^\dagger_i c^\dagger_j c_k c_l+ \frac{1}{\sqrt{N}}\sum_{ij} J_{ij} c^{\dagger}_i c_j \label{eq:SYK2_SYK4}
\end{equation}
where the couplings $J_{ijkl}, J_{ij}$ are same as in the SYK$_4$ and SYK$_2$ model, respectively. The model has a Fermi liquid (FL) ground state with Landau quasiparticle excitations. The model is a slow scrambler \cite{BAmodel2017,Kim2020} and slow thermalizer \cite{Eberlein2017} with the Lyapunov exponent and thermalization rate both proportional to $T^2$, set by quasiparticle scattering at low temperature $T\ll J_2^2/J_4$ ($J_2$) for $J_2 \ll J_4$ ($\gg J_4$).  


Finally, we consider the $(q/2)$-body ($q$ even) generalization of the SYK$_4$ model whose Hamiltonian is given by 
\begin{equation} \label{SYKqHam}
    \mathcal{H}_q = \sum_{\substack{i_1, \cdots, i_{q/2} \\ j_1, \cdots j_{q/2}}} J_{i_1 \cdots i_{q/2}; j_1 \cdots j_{q/2}} c^\dagger_{i_{q/2}} \cdots c^\dagger_{i_{1}} c_{j_1} \cdots c_{j_{q/2}} 
\end{equation}
where the couplings $J_{i_1 \cdots i_{q/2}; j_1 \cdots j_{q/2}}$ are independent Gaussian random numbers with zero mean and the variance $\expval{|J_{i_1, \dots, i_{q/2}; j_1, \dots j_{q/2}}|^2} =  J_q^2/[(q/2)!^2 N^{q-1} (q/2)]$, while satisfying appropriate anti-symmetry and Hermiticity relations. In the large-$q$ limit, this model is known to be a \emph{perfect thermalizer}, leading to instantaneous thermalization of mixed states after a quench \cite{Eberlein2017, Louw_manySYK}.

\section{Schwinger-Keldysh field theory for time evolution of pure states} \label{SKPureStateFormalism}
Schwinger-Keldysh (SK) field theory \cite{Kamenev2011,Stefanucci2013} is a general framework to study non-equilibrium dynamics of quantum many-body systems. In the SK framework, the dynamics is described in terms of the generating function $Z=\mathrm{Tr}[\rho(t)]=\mathrm{Tr}[U(\infty,0)\rho(0)U(0,\infty)]$ which is used to obtain the time-dependent correlation functions. Here, $\rho(0)$ is the initial density matrix and $U(\infty,0)=U(t\to \infty,0)$ is the unitary time evolution operator for a Hamiltonian $\mathcal{H}(t)$, which can be time dependent in general. As discussed in the introduction, here we consider an initial pure-state density matrix and write the generating function as a coherent-state path integral on a SK contour $\mathcal{C}$, comprising a forward ($+$) and a backward ($-$) time branch [Fig.\ref{fig:KeldyshContour}], to obtain
\begin{equation}
Z = \int \mathcal{D}(\bar{c}, c) e^{\ci S} \bra{c(0+)} \rho(0)\ket{-c(0-)}. \label{eq:SK_Z}
\end{equation}
Here, $S = \int_{\mathcal{C}} dz [\sum_i \bar{c}_i(z) \ci \partial_z c_i(z) - \mathcal{H}(z,\bar{c}(z), c(z))]$ is the action, and $\bar{c}_i(z),~c_i(z)$ are fermionic Grassmann fields on the SK contour $z=(t\pm)\in \mathcal{C}$. $\mathcal{H}(z,\bar{c}(z), c(z))=\mathcal{H}(t,\bar{c}(z),c(z))$ is the coherent-state matrix element of normal-ordered Hamiltonian. To proceed further and calculate correlation functions from the SK path integral, it is crucial to exponentiate the matrix element $\bra{c(0+)} \rho(0)\ket{-c(0-)}$ and incorporate it into the action, as we discuss below.

The standard SK approach \cite{Kamenev2011} incorporates a density matrix $\rho(0)$ at $t=0$ assuming an adiabatic evolution, starting from a density matrix $\rho(-\infty)$ of a suitable non-interacting Hamiltonian $\mathcal{H}_0$ in the distant past $t\to -\infty$. Here the interaction is slowly switched on from $t\to -\infty$ to $t=0$ to arrive at the desired Hamiltonian $\mathcal{H}(t)$ for $t>0$. The density matrix $\rho(-\infty)\propto e^{-\beta \mathcal{H}_0}$ describes thermal state of the non-interacting Hamiltonian at an inverse temperature $\beta=1/T$ ($k_\mathrm{B}=1$), or the ground state for $T\to 0$. The system is assumed to thermalize over the interval $-\infty <t\leq 0$ and reach the thermal state $\rho(0)\propto e^{-\beta \mathcal{H}(0^-)}$ of the interacting Hamiltonian at $t=0$. Further evolution for $t>0$, for examples, after a quench $\mathcal{H}(0^-)\to \mathcal{H}(0^+)\neq \mathcal{H}(0^-)$ at $t=0$ or under a general time dependent Hamiltonian $\mathcal{H}(t>0)$, can then be described by the SK field theory. However, adiabatic assumption may not always hold, particularly if one is trying to reach from one pure sate to other, i.e., from the ground state of $\mathcal{H}_0$ to that of $\mathcal{H}(0^-)$. More importantly, one needs to design Hamiltonians $\mathcal{H}_0$ and $\mathcal{H}$ to arrive at a desired pure state. It is not clear whether this can be done for an arbitrary pure-state density matrix $\rho(0)=|\Psi(0)\rangle\langle \Psi(0)|$ with $|\Psi(0)\rangle =\sum_n C_n |n\rangle$. Also, the standard SK approach \cite{Kamenev2011} is often used for steady state, even in non-equilibrium situations, where all initial correlations are neglected.


As an alternative to the above adiabatic approach, the Konstantinov-Perel (KP) contour \cite{Stefanucci2013} is useful when the initial density matrix at $t=0$ can be written as an exponential of a many-body operator, i.e., $\rho(0) \sim \exp(-\beta \mathcal{H}^M) \sim \exp(\int_0^{-\ci \beta} dz \mathcal{H}^M)$, for some positive constant $\beta$. In this case, a vertical branch from $0$ to $0-\ci \beta$ is added to the SK contour, yielding the full KP contour $[0, \infty) \cup (\infty, 0] \cup [0, 0-\ci \beta]$. This method is useful to study quantum quenches starting from a thermal density matrix of a Hamiltonian $\mathcal{H}_0$ and time evolving with a different Hamiltonian $\mathcal{H}(t>0)$. However, for a generic pure state, it is difficult to find such a convenient Hamiltonian $\mathcal{H}^M$, particularly a local one. More recently, Chakraborty et al. \cite{Chakraborty2019,Chakraborty2020} have proposed an extension of the SK method to incorporate arbitrary initial density matrices. However, as discussed in Sec.\ref{sec:Overview}, this method is not always easy to apply to interacting models, like the SYK models considered here. Below, we formulate a general SK method that can be applied to describe evolution of pure states in generic interacting systems, including the strongly interacting SYK models.


\subsection{Pure product state}
We first elucidate the basic steps of the new SK method for studying the time evolution of a pure product state in the Fock space, 
\begin{equation}
    \ket{\Psi(0)} = \ket{n} = \ket{n_1, \dots, n_N} \label{eq:FockState}.
\end{equation}
with $i=1,2,\cdots, N$ sites. Here $n_{i} = 1$ for $i\in I$ and $n_{i} = 0$ for $i\in \bar{I}$, where $I$ is the set of filled or occupied $N_I = fN$ sites and $\bar{I}$ is the set of empty sites for filling $f=(1/N)\sum_i n_i$ . 
For a fermionic coherent state $\ket{-\psi} \equiv \exp(-\sum_i c^\dagger_i \psi_i) \ket{0}$, we have $\braket{n}{-\psi} = \psi_N^{n_N} \dots \psi_1^{n_1}$, and similarly $\langle \psi|n\rangle=(-\bar{\psi}_1)^{n_1}\cdots (-\bar{\psi}_N)^{n_N}$. Using this, we obtain the matrix element of the initial density matrix appearing in the SK path integral of Eq.\eqref{eq:SK_Z} 
\begin{equation}
\braket{c(0+)}{n} \braket{n}{-c(0-)} = \prod_{i \in I} (-\bar{c}_i(0+)c_i(0-)), 
\end{equation}
where only the terms corresponding to initially occupied sites appear in the product over the set $I$. The key identity used to exponentiate this term is $(-\bar{\psi} \psi) = \int d \bar{\eta} d \eta \exp(\bar{\psi} \eta - \bar{\eta} \psi) $ with Grassmann variables $\bar{\eta},\eta$. Thus, we get
\begin{multline}
 \braket{c(0+)}{n} \braket{n}{-c(0-)} \\= \int \prod_{i \in I} d \bar{\eta}_i d \eta_i \exp(\sum_{i \in I} \bar{c}_i(0+)\eta_i - \bar{\eta}_i c_i(0-)) ,
\end{multline}
where we have introduced the \emph{auxiliary} static Grassmann source fields $\{\bar{\eta}_i,\eta_i\}_{i \in I}$ to keep track of the initially occupied sites. As a result, we write the SK generating function of Eq.\eqref{eq:SK_Z} as
\begin{equation}
Z = \int \mathcal{D}(\bar{c}, c) \prod_{i \in I}d \bar{\eta}_i d \eta_i \ e^{\ci(S + S_{in})}. \label{eq:SK_Z_SYKFock}
\end{equation}
with an additional term $S_{in}$ that encodes the initial pure state in the action. Here
\begin{subequations} \label{eq:SKAction_FockState}
\begin{align}
S &= \int_{\mathcal{C}} dz \left[\sum_i \bar{c}_i(z) \ci \partial_z c_i(z) - \mathcal{H}(z,\bar{c}(z),c(z))\right] \label{SKactionH}\\
S_{in} &= -\ci \int_{\mathcal{C}} dz \sum_{i \in I} [\bar{c}_i(z)\delta_{\mathcal{C}}(z,0+)\eta_i - \bar{\eta}_i \delta_{\mathcal{C}}(z,0-)c_i(z) ] \label{eq:Sin},
\end{align} 
\end{subequations}
with Dirac delta function $\delta_\mathcal{C}(z,z')$ on the SK contour.
The auxiliary static fields $\{\bar{\eta}_i,\eta_i\}_{i\in I}$ are integrated out in the SK path integral along with the dynamical fermionic fields $\{\bar{c}_i(z),c_i(z)\}$. We generalize the method below for an arbitrary pure state in a fixed particle number sector.

\subsection{General pure state in a fixed particle number sector}\label{sec:SK_ArbPure}
Here we consider an initial state
\begin{equation}
    \ket{\Psi(0)} = \sum_n C_n \ket{n},\label{eq:genFockState}
\end{equation} 
which is superposition of different Fock states $|n\rangle$ with $(1/N)\sum_i n_i=f$ for all the $|n\rangle$ appearing above. As above, we denote the set of occupied (empty) sites in $|n\rangle$ as $I_n$ ($\bar{I}_n$). Thus the initial density matrix in Eq.\eqref{eq:SK_Z} is $\rho(0)=\sum_{nn'}C_nC_{n'}^*|n\rangle \langle n'|$. We proceed as in the preceding section to exponentiate the density matrix element using
\begin{align}
    &\braket{c(0+)}{n} \braket{n'}{-c(0-)}  \nonumber \\
    &= \prod_{i \in I_n, i' \in I_{n'}} (-\bar{c}_i(0 +) c_{i'}(0-)) \nonumber \\
    &= \int \prod_{\substack{i \in I_n\\i' \in I_{n'}}} d\bar{\eta}_{i'}d\eta_{i} \exp(\sum_{i \in I_n} \bar{c}_i(0,+)\eta_i - \sum_{i' \in I_{n'}} \bar{\eta}_{i'} c_{i'}(0,-)). 
\end{align}
In the above we have ordered and paired up the terms according to $i_1<\cdots<i_{fN}\in I_n$ and $i'_1<\cdots<i'_{fN}\in I_{n'}$.
The SK generating function is then given by 
\begin{equation}
        Z = \int   \mathcal{D}(\bar{c}, c) e^{\ci S} \sum_{nn'} C_n C^*_{n'} \prod_{i \in I_n, i' \in I_{n'}} d\bar{\eta}_{i'}d\eta_{i}  e^{\ci S_{nn'}}, \label{eq:SK_Z_ArbPure}
\end{equation}
with the standard SK action $S$, as in Eq.\eqref{SKactionH}, and 
\begin{equation}
    \ci S_{nn'} =   \int_{\mathcal{C}} dz \sum_{i \in I_n } \bar{c}_i(z)\delta_{\mathcal{C}}(z,0+)\eta_i - \sum_{i' \in I_{n'} } \bar{\eta}_{i'} \delta_{\mathcal{C}}(z,0-)c_{i'}(z). \label{eq:Snn'}
\end{equation}

\begin{figure*}[ht]
    \centering
    \includegraphics[width=\textwidth]{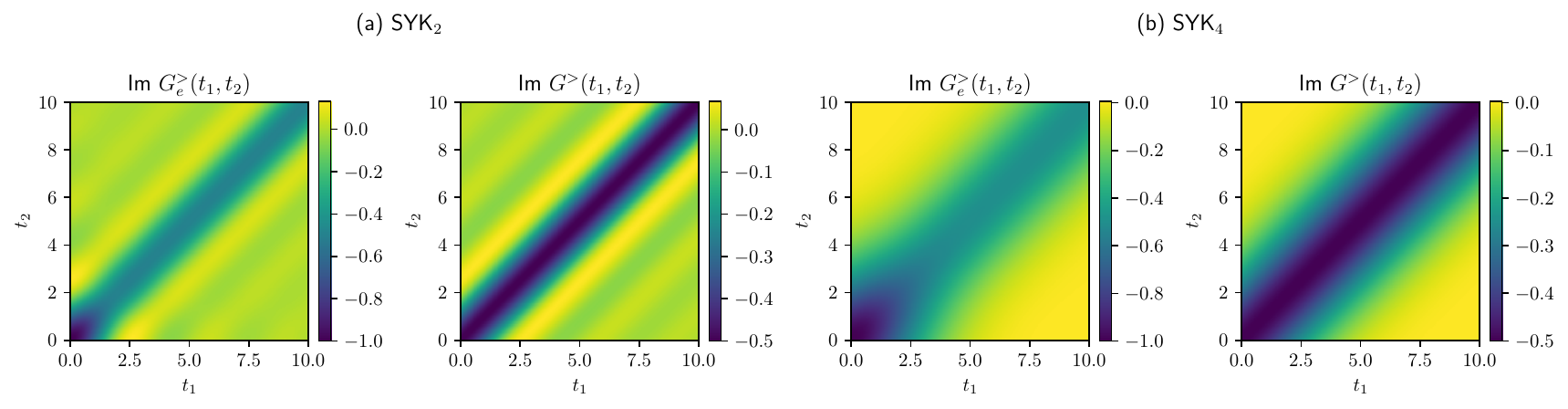}
    \caption{{\bf Time evolution of the non-equilibrium Green's function for SYK$_2$ and SYK$_4$ models:} Green's function of initially unfilled sites, $G^>_e(t_1, t_2)$, and the large-$N$ collective Green's function, $G^>(t_1, t_2)$ for initial pure Fock state in the (a) SYK$_2$ and (b) SYK$_4$ models at half-filling, $f=0.5$. From the Kadanoff-Baym equations [Eqs.\eqref{KBEqns1}] and the initial Green's functions at $(0,0)$ [Eqs.\eqref{eq:InitialCondition_Ge}], it is easy to see that these Green's functions are purely imaginary.}
    \label{fig:SYKGreensFn}
\end{figure*}

\section{Large-$N$ SK field theory for evolution of pure states in SYK models}\label{SYKlargeN}
\subsection{Initial unentangled pure product state}
We use the formalism of Section \ref{SKPureStateFormalism} above to study the non-equilibrium dynamics of pure states in the SYK model. We first consider the time evolution of pure product state of Eq.\eqref{eq:FockState} and the corresponding SK generating function [Eq.\eqref{eq:SK_Z_SYKFock}] and action [Eq.\eqref{eq:SKAction_FockState}]. As discussed in the Appendix \ref{SK_actionKBEqns} in detail, we obtain the effective SK action for generating disordered-averaged Green's functions $G_{ij}(z_1,z_2)=-\ci \overline{\langle c_i(z_1)\bar{c}_j(z_2)\rangle}$ by averaging ($\overline{\langle \cdots\rangle}$) the non-equilibrium generating function $Z$ [Eq.\eqref{eq:SK_Z_SYKFock}] over all disorder realizations of random $\mathrm{SYK}_q$ couplings $J_{i_1\cdots i_{q/2},j_1\cdots j_{q/2}}$. Introducing the large-$N$ collective Green's function $G(z_1, z_2) = -(\ci/N) \sum_{i}c_{i}(z_1)\bar{c}_{i}(z_2) $ and its conjugate $\Sigma(z_1, z_2)$ in the resultant generating function, we integrate out the fermionic fields to obtain the non-equilibrium $G-\Sigma$ action for the pure state evolution. In the large-$N$ limit, we obtain the following closed set of saddle point equations from the action
\begin{subequations} \label{eq:Saddle_Fock}
\begin{align}
G_{e}^{-1}(z_1,z_2) &= \mathrm{i} \partial_{z_1} \delta(z_1-z_2) - \Sigma(z_1,z_2)\\
G_{ii}(z_1,z_2) &= G_{e}(z_1,z_2) -\delta_{i\in I} \frac{G_{e}(z_1,0+)G_{e}(0-,z_2)}{G_{e}(0-,0+)}\label{Gi}\\
  G(z_1, z_2) &= \frac{1}{N} \sum_{i=1}^N G_{ii}(z_1, z_2) \label{eq:G}\\
  \Sigma(z_1, z_2) &= J_q^{2}G(z_1,z_2)^{q/2}G(z_2,z_1)^{q/2-1} \label{eq:Sigma}
\end{align}
\end{subequations}
Here $\delta_{i\in I}=1$ if $i\in I$ and $\delta_{i\in I}=0$ if $i\in \bar{I}$.
The large-$N$ theory yields two types of Green's functions, as inferred from Eq.\eqref{Gi} -- one corresponding to initially filled sites, which we define as $G_f(z_1, z_2)=G_e(z_1,z_2)-G_e(z_1,0+)G_e(0-,z_2)/G_e(0-,0+)$, and another corresponding to initially unfilled sites, which we define as $G_e(z_1, z_2)$. $G_f$ is determined by $G_e$ through Eq.\eqref{Gi}. As a result, the large-$N$ Green's function,
\begin{align}
G(z_1,z_2) &= fG_{f}(z_1,z_2)+(1-f)G_{e}(z_1,z_2), \label{eq:LargeN_G}
\end{align}
is also determined by $G_e$.
The self-energy $\Sigma(z_1,z_2)$ to determine $G_e(z_1,z_2)$ is obtained from the large-$N$ Green's function from Eq.\eqref{eq:Sigma}. Thus the Eqs.\eqref{eq:Saddle_Fock} form a closed set of self-consistency equations. In Sec.\ref{Results}, we solve these self-consistency equations numerically. \redtext{Moreover, we obtain exact analytical solutions of Eqs.\eqref{eq:Saddle_Fock} for the $\mathrm{SYK}_2$ [Eq.\eqref{eq:SYK2}] and large-$q$ $\mathrm{SYK}_q$ [Eq.\eqref{SYKqHam}] models. We also obtain analytical expression for early and asymptotic late-time relaxation of initial density inhomogeneity for the $\mathrm{SYK}_q$ model [Eq.\eqref{eq:SYK4}] for general $q\geq 4$.}

\subsection{Initial entangled states} \label{PureEntStateResults}
Using the formalism of Sec.\ref{sec:SK_ArbPure} and following steps similar to the case of pure product state above, we also obtain the large-$N$ saddle-point equations for a general initial pure state $|\Psi(0)\rangle=\sum_n C_n |n\rangle$ ($\sum_n |C_n|^2=1$) of Eq.\eqref{eq:genFockState}. Here, while disorder averaging the SK generating function, it is important to note that $C_n$'s are uncorrelated with random SYK couplings. Thus, the pure initial state $|\Psi(0)\rangle$ considered here exclude eigenstates of the SYK Hamiltonian. As discussed in detail in Appendix \ref{app:ArbPure_SYK}, the large-$N$ saddle-point equations, e.g., for SYK$_q$ model, turn out to be same as Eqs.\eqref{eq:Saddle_Fock}, except the expression for $G_{ii}(z_1,z_2)$ in Eq.\eqref{Gi} is replaced by
\begin{align}
G_{ii}(z_1,z_2)&=G_e(z_1,z_2)\nonumber \\
&-\frac{G_{e}(z_1,0+)G_{e}(0-,z_2)}{G_{e}(0-,0+)}\sum_n|C_n|^2\delta_{i\in I_n}. \label{eq:LocalG_ArbPure}
\end{align}
However, the above leads to the same expression of Eq.\eqref{eq:LargeN_G} for the large-$N$ Green's function, as well as the initial conditions [Eqs.\eqref{eq:InitialCondition_Ge}] for $G_e$ or $G$. Thus, even for an arbitrary initial pure state $|\Psi(0)\rangle=\sum_n C_n |n\rangle$, the large-$N$ self-consistency equations remain exactly the same as that for a pure product Fock state. Remarkably, this implies that the non-equilibrium dynamics of all initial generic pure states in a fixed particle number sector for a particular SYK model or a combinations of SYK models, like SYK$_q$ or SYK$_2$+SYK$_4$, is entirely captured by a single universal Green's function, $G_e(z_1,z_2)$. The latter is completely determined by a universal set of self-consistency equations, e.g., Eqs.\eqref{eq:SelfConsistency_SYK} of Appendix \ref{SK_actionKBEqns} for SYK$_q$, along with the initial condition of Eq.\eqref{eq:Ge_Initial}. The large-$N$ self-consistent saddle-point equations only depend on the average fermion density or filling $f$. Of course, the equation connecting the self-energy $\Sigma(z_1,z_2)$ and the large-$N$ Green's function $G(z_1,z_2)$, such as Eq.\eqref{eq:SelfEnergy_SYKq_A}, depends on the particular type of SYK model or the combinations of the SYK models.

Nevertheless, different initial pure states, in general, will exhibit distinct non-local and/or multi-point correlations with non-trivial time evolutions, albeit all determined by various combinations of a single universal Green's function $G_e(z_1,z_2)$, as discussed in Appendix \ref{app:ArbPure_SYK}. For example, a general two-point correlation function or Green's function is given by 
\begin{align}
&G_{ij}(z_1,z_2)=-\ci\langle c_i(z_1)\bar{c}_j(z_2)\rangle_c\nonumber \\
&=\delta_{ij}\sum_n |C_n|^2 G^{(n)}_{ii}(z_1,z_2)+\ci \sum_{\substack{n\neq n'\\d_{nn'}N=2}}\frac{C_nC_{n'}^*}{[\ci G_e(0-,0+)]^{2f}}\nonumber \\
&\times \delta_{i\in I_n\cap \bar{I}_{n'}}\delta_{j\in \bar{I}_n \cap I_{n'}}G_e(z_1,0+)G_e(0-,z_2). \label{eq:Gij_ArbPure}
\end{align}
In the above, $d_{nn'}N$ is the hamming distance between the Fock states $|n\rangle$ and $|n'\rangle$. Any higher-point correlation functions can also be written solely in terms of $G_e(z_1,z_2)$ (Appendix \ref{app:ArbPure_SYK}).

To demonstrate application of our results for initial entangled states, we first consider an initial state with a Bell pair between site 1 and 2 (arbitrary),
\begin{align}
|\Psi(0)\rangle =\frac{1}{\sqrt{2}}\left(|10n_3\cdots n_N\rangle+|01n_3\cdots n_N\rangle \right) \label{eq:BellPairState}
\end{align}
Using the general expression of Eq.\eqref{eq:Gij_ArbPure}, it is easy to show that the only non-zero two-point functions are
\begin{subequations}
\begin{align}
G_{11}(z_1,z_2)&=G_{22}(z_1,z_2) \nonumber \\
&=G_e(z_1,z_2)-\frac{\ci}{2}G_e(z_1,0+)G_e(0-,z_2) \label{eq:BellPair_localG}\\
G_{ii}(z_1,z_2)&|_{i\neq 1,2}=G_e(z_1,z_2)-\ci n_i G_e(z_1,0+)G_e(0-,z_2) \label{eq:BellPair_localGi}\\
G_{12}(z_1,z_2)&=G_{21}(z_1,z_2)=\frac{\ci}{2}G_e(z_1,0+)G_e(0-,z_2), \label{eq:BellPair_Corrlation}
\end{align}
\end{subequations}
where we have used the initial condition [Eq.\eqref{eq:Ge_Initial}] $G_e(0-,0+)=G^>_e(0,0)=-\ci$. We see that non-local correlations $G_{12},G_{21}$ exist between initially entangled sites. 

Next, we consider a Greenberger–Horne–Zeilinger (GHZ)-type initial state at half filling $f=1/2$,
\begin{align}
|\Psi(0)\rangle =\frac{1}{\sqrt{2}}\left(|1\rangle +|2\rangle\right)=\frac{1}{\sqrt{2}}\left(|n_1\cdots n_N\rangle +|\bar{n}_1\cdots \bar{n}_N\rangle\right),
\end{align}
where $\bar{n}_i=0 (1)$ for $n_i=1 (0)$, with $N$ even. In this case, the local Green's function is $G_{ii}(z_1,z_2)=G_e(z_1,z_2)-(\ci/2)G_e(z_1,0+)G_e(0-,z_2)$ for all $i$. However, there exists a non-trivial $N$-point function in this case, namely for $i_1<\cdots<i_{N/2}\in I_1;j_1<\cdots<j_{N/2}\in I_2$,
\begin{align}
&G_{i_1\cdots i_{N/2}j_1\cdots j_{N/2}}(z_1,\cdots,z_{N/2},z_1',\cdots,z'_{N/2}) \nonumber \\
&=\frac{1}{\ci^{N/2}}\langle c_{i_1}(z_1)\cdots c_{i_{N/2}}(z_{N/2})\bar{c}_{j_{N/2}}(z'_{N/2}) \cdots \bar{c}_{j_1}(z_1')\rangle \nonumber \\
&=\frac{1}{2}G_e(z_1,0+)\cdots G_e(z_{N/2},0+) G_e(0-,z_1')\cdots G_e(0-,z_{N/2}')
\end{align}

The above examples illustrate how the entanglement of the initial pure state manifests itself as non-local and/or multi-point correlations in the non-equilibrium dynamics. In the SYK models, however, all these correlations are entirely determined by a universal two-point Green's function $G_e(z_1,z_2)$. Since, the Green's functions $G_e(ts,0+)$, $G_e(0-,ts)$ ($s=\pm$) decay with time $t$, as we show in the next section, all the non-local and multi-point quantum correlations due to the entanglement in the initial state decay as the system thermalizes. Eventually only the local two-point Green's function $G_{ii}(t_1s,t_2s')=G_{ii}^{ss'}(t-t')$ survives in the steady state, as expected for the large-$N$ SYK models.

\section{Results}\label{Results}

\subsection{Time evolution of pure Fock state in $\mathrm{SYK}_2$ and $\mathrm{SYK}_4$ models}\label{PureFockStateResults}
The saddle-point Eqs.\eqref{eq:Saddle_Fock} can be solved efficiently by casting them as causal Kadanoff-Baym (KB) equations. In Appendix \ref{SK_actionKBEqns}, we derive the KB equations for the disorder-averaged non-equilibrium Green's functions, $G_{e}^>(t_1, t_2) =G_{e}(t_1-,t_2+)$ and $G_{e}^<(t_1, t_2) = G_{e}(t_1+,t_2-)$. The KB Eqs.\eqref{KBEqns1} are solved using a predictor-corrector scheme on a discretized $t_1-t_2$ grid $[0, T_{\text{max}}] \times [0, T_{\text{max}}]$ with a step size of $\delta t=0.05$ along both directions (see Supplemental Material of Ref.\cite{Arijit2020} for details). The initial condition for the pure product state $|n\rangle$ is given by $G_e^>(0,0)=-\ci$ and $G_e^<(0,0)=0$. The integrals are evaluated using the trapezoidal rule. We ensure the consistency of the numerical integration by monitoring the conservation of energy and average fermion density, as well as verifying convergence with $T_{\text{max}}$, step size, and number of corrector steps.

In both SYK$_2$ and SYK$_4$ models, we find that the energy of the pure Fock state $|n\rangle$ is zero. This implies that the pure state belongs to the infinite-temperature state at the middle of the many-body energy spectrum \cite{ChowdhurySYKReview} of the SYK model. We find that the large-$N$ Green's function depends only on the relative time, i.e., $G^{>,<}(t_1, t_2) = G^{>,<}(t_1 - t_2)$ for all $t_1, t_2$, as shown in Fig.\ref{fig:SYKGreensFn}. To track the effective temperature as a function of time, we define the Wigner transform of the Green’s function $G(\mathcal{T}, \omega) = \int_{-\infty}^{\infty} dt e^{\ci \omega t} G(\mathcal{T} + t/2, \mathcal{T} - t/2)$ and the occupation function $f(\mathcal{T}, \omega) = \ci G^<(\mathcal{T}, \omega)/2 \Im G^R(\mathcal{T}, \omega)$, which approaches the Fermi function when the system thermalizes \cite{Bruus2004-py}. Here $G^R(t_1,t_2)=\theta(t_1-t_2)[G^>(t_1,t_2)-G^<(t_1,t_2)]$ is the retarded Green's function. We fit the occupation function at low frequencies with a Fermi function and obtain the effective temperature. We find that the effective temperature of the pure Fock state is infinite (not shown). The effective temperature extracted from large-$N$ Green's function, of course, does not evolve with time since $G(t_1, t_2)$ instantaneously thermalizes. Interestingly, even though the large-$N$ Green’s function $G(t_1, t_2)$ instantaneously thermalizes to infinite temperature, the local Green’s functions $G_{e/f}(t_1,t_2)$ relax at a finite rate, as seen for $G_e$ in Fig.\ref{fig:SYKGreensFn}. 

\redtext{The instantaneous thermalization of the large-$N$ Green's function is not only limited to initial pure Fock state $|n\rangle$, but holds for any generic initial state $|\Psi(0)\rangle=\sum_n C_n |n\rangle$. The dynamics of the latter is also universally described by the same KB Eqs.\eqref{KBEqns1}, the initial conditions [Eqs.\eqref{eq:InitialCondition_Ge}] and $G_e(t_1,t_2)$, as discussed in the preceding section. Just like $|n\rangle$, any generic state $|\Psi(0)\rangle$ also corresponds to infinite-temperature state with respect to random SYK Hamiltonians.}

\redtext{In Appendix \ref{app:G_KB_Thermalization}, we provide an analytical proof of the instantaneous thermalization of the large-$N$ Green's function $G(t_1,t_2)$ under the dynamics described by the KB Eqs.\eqref{KBEqns1} and the initial conditions [Eqs.\eqref{eq:InitialCondition_Ge}]. Moreover, based on many-body random matrix theory in Sec.\ref{sec:G_RMT_Thermalization}, we show that the instantaneous thermalization of the disorder and site-averaged local Green's function $G(t_1,t_2)$, starting from generic pure state, is a general property of any random Hamiltonian in the thermodynamic limit.  We also obtain closed-form analytical expressions for $G$ and $G_e$ in the non-interacting SYK$_2$ model by exactly solving the corresponding KB equations [Eqs.\eqref{KBEqns1},\eqref{eq:InitialCondition_Ge}] and by using single-particle RMT in Sec.\ref{sec:RMT_singleparticle}. The analytical expressions exactly matches (not shown) with the large-$N$ numerical results, e.g., shown in Fig.\ref{fig:SYKGreensFn}(a). Thus, the analytical results for $\mathrm{SYK}_2$ also confirms the instantaneous thermalization of the large-$N$ Green's function. The instantaneous thermalization of $G(t_1,t_2)$ implies $\mathcal{T}$ independence  of site and disorder averaged fermionic occupation, $f(\mathcal{T},\omega)=f(\omega)$, and density of states, $ -(1/\pi)\mathrm{Im}G^R(\mathcal{T},\omega)=-(1/\pi)\mathrm{Im}G^R(\omega)$, which are physical observables.}

\begin{figure}[ht]
        \centering
        \includegraphics[width=\linewidth]{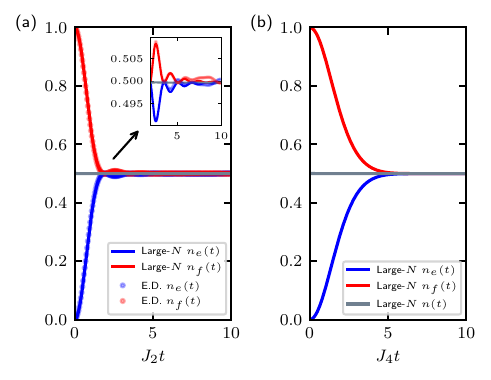}
        \caption{{\bf Relaxation of density inhomogeneity of the half-filled pure Fock state in SYK$_2$ and SYK$_4$ models:} Time evolution of the density of initially empty sites $n_e(t)$ and filled sites $n_f(t)$ of a pure Fock state at half-filling ($f = 1/2$) in the (a) SYK$_2$ and (b) SYK$_4$ models. The inset in panel (a) shows a magnified plot of the decaying oscillations in the SYK$_2$ model. The large-$N$ predictions for the SYK$_2$ model are in excellent agreement with exact diagonalization results for a system size $N=64$ and $200$ disorder realizations, for $J_2 t \lesssim 5$.}
        \label{fig:SYKHalfFilling}
    \end{figure}

Next, we investigate the fate of initial density inhomogeneity or imbalance between initially filled and unfilled sites, namely how the inhomogeneity relaxes under non-equilibrium dynamics starting from a pure state and the system becomes homogeneous. To this end, we calculate the densities of the initially filled and unfilled sites using $n_{e,f}(t) = - \ci G_{e,f}^<(t,t)$. First, we start with a half-filled state $|n\rangle$ in Fig.\ref{fig:SYKHalfFilling}. In this case, density at all the sites should approach 1/2 when the system thermalizes. As shown in Fig.\ref{fig:SYKHalfFilling}(a), the densities $n_{e,f}(t)$, obtained from the large-$N$ saddle point [Eqs.\eqref{eq:Saddle_Fock}], have an oscillatory relaxation towards the steady-state value for the non-interacting $\mathrm{SYK}_2$ model [Eq.\eqref{eq:SYK2}]. The densities decays to a small value within a time $t\sim J_2^{-1}$ and then continue to oscillate with a decaying amplitude. To benchmark our large-$N$ results, we compute $n_{e,f}(t)$ for large but finite $N$ ($N=64$) by numerically diagonalizing the $N\times N$ random matrix $J_{ij}$ in Eq.\eqref{eq:SYK2} and obtain the disorder-averaged densities by averaging over many realizations of $\{J_{ij}\}$ (see Appendix \ref{ED_SYK2}). We observe an excellent agreement with the large-$N$ KB calculations in Fig.\ref{fig:SYKHalfFilling}(a) up to $J_2 t \lesssim 5$. For later times, one needs more disorder averaging or larger system sizes for agreement with the large-$N$ results. The exact agreement between large-$N$ results and finite-$N$ exact diagonalization results substantiate our large-$N$ SK formalism for pure state.

\begin{figure}[ht]
    \centering
    \includegraphics[width=\linewidth]{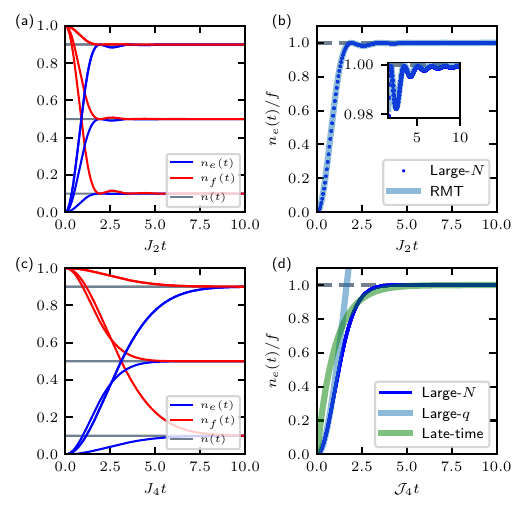}
    \caption{{\bf Density dependent thermalization of pure Fock state in SYK$_2$ and SYK$_4$ models:} Dependence of the non-equilibrium dynamics of density inhomogeneity in the initial pure state on the fermion filling ($f$) in the (a) SYK$_2$ and (c) SYK$_4$ models. Results for three values of filling, $f=0.1,0.5,0.9$, are shown. (b) Scaling collapse of the dynamics of initially unfilled sites in the SYK$_2$ model: $n_e(t)/f = f_2(J_2t)$, where the scaling function $f_2(t)$, known from \redtext{analytical solution of Kadanoff-Baym equations [Sec.\ref{sec:KBAnalytic_SYK2}]} and RMT result of Eq.\eqref{RMTnc}, shows exact agreement with the large-$N$ results obtained by numerically solving the Kadanoff-Baym equations. The inset shows a magnified plot of the decaying oscillations. (d) Scaling collapse in the SYK$_4$ model: $n_e(t)/f = f_4(\mathcal{J}_4t)$, where $\mathcal{J}_4 = J_4 \sqrt{2f(1-f)}$ is the density-dependent emergent inverse time scale. The large-$q$ result of Eq.\eqref{Largeqnc} with $q=4$ agrees well in the regime $\mathcal{J}_4 t \lesssim O(1)$. \redtext{The green curve represents the late-time asymptotic analytical result Eq.\eqref{eq:ne_latetime_SYK4}, shows agreement in the regime $\mathcal{J}_4 t \gtrsim O(2.5) $.} }
    \label{fig:SYKDensityDependence}
\end{figure}

In contrast to the oscillatory relaxation of the densities in the $\mathrm{SYK}_2$ model, the large-$N$ dynamics in the $\mathrm{SYK}_4$ model exhibits monotonic relaxation to the steady state, as shown in Fig.\ref{fig:SYKHalfFilling}(b). Curiously, the decay of the inhomogeneity is slower ($t\sim 5 J_4^{-1}$) for the interacting $\mathrm{SYK}_4$ model compared to dominant initial decay ($t\sim J_2^{-1}$) in non-interacting $\mathrm{SYK}_2$ model. The slower thermalization in the interacting SYK$_4$ model compared to non-interacting SYK$_2$ model for infinite-temperature pure states is in contrast to the thermalization of low-temperature mixed states. For the latter, the weakly interacting Fermi liquid thermalizes slower compared to the strongly interacting non-Fermi liquid regimes of the SYK models \cite{Eberlein2017,Arijit2020}.

\redtext{We obtain exact solutions for $G^{>,<}(t_1,t_2)$, $G_e^{>,<}(t_1,t_2)$ and $n_{e,f}(t)$ from the KB equations [Eqs.\eqref{KBEqns1},\eqref{eq:InitialCondition_Ge}] for $\mathrm{SYK}_2$ model below. In the next section, the same expressions are obtained from single-particle RMT, which shows that the relaxation in the SYK$_2$ model is due to the dephasing of random phases. Since the SYK$_2$ is a single-particle random matrix model, it is analytically tractable using RMT. We also obtain an estimate of the thermalization rate of the imbalance for the pure state in the SYK$_{q\geq 4}$ model from late-time asymptotic solutions of the KB equations [Eqs.\eqref{KBEqns1},\eqref{eq:InitialCondition_Ge}] and from a large-$q$ approximation. The $(q/2)$-body generalization of the SYK$_4$ model in the large-$q$ limit offers analytic tractability to the interacting model \cite{SYK2016}. These analyses show that the rate depends on the filling $f$.} The filling dependence of the rate in SYK$_4$ model makes it slower compared to the SYK$_2$ model, where the rate only depends on $J_2$. The filling-dependent relaxation rate at infinite temperature for the SYK$_4$ model can be understood heuristically from phase space or kinematic constraints on the four-fermion interaction of Eq.\eqref{eq:SYK4} to relax density inhomogeneity. The four-fermion interaction involves a pair of fermions \emph{hopping} from two filled sites to two empty sites.

The dependence of the dynamics on the filling and the lack thereof for the SYK$_4$ and SYK$_2$ models, as obtained from the numerical solutions of large-$N$ KB equations, are shown in Figs.\ref{fig:SYKDensityDependence} (a) and (c). \redtext{As evident, the time scale for relaxation of the inhomogeneity is independent of the filling $f$ in the SYK$_2$ model}. On the contrary, in the SYK$_4$ model, the time scale exhibits density dependence, becoming slower as the system moves away from half-filling. Below, we gain analytical insights into the contrast between the dynamics in interacting and non-interacting models. 

\subsection{\redtext{Analytical solutions of Kadanoff-Baym equations for the time evolution of pure states}}\label{sec:KB_simplification}

\redtext{The instantaneous thermalization of the large-$N$ Green's function, i.e., $G^{>,<}(t_1, t_2) = G^{>,<}(t_1-t_2)$, allows us to significantly simplify the KB Eqs.\eqref{KBEqns1}. Since the self-energies are products of large-$N$ Green's functions, they also instantaneously thermalize, $\Sigma^{>,<}(t_1, t_2) = \Sigma^{>,<}(t_1-t_2)$. From Eqs.\eqref{KB2}\eqref{KB3}, we can invert the relation between $G_e, G$ by plugging in $t_1 = 0$ or $t_2 = 0$ to get 
\begin{equation}\label{Eq:Ge}
    G_e^{>,<}(t_1, t_2) = G^{>,<}(t_1-t_2) +  \frac{f}{(1-f)^2} \frac{G^>(t_1) G^>(t_2)}{G_e^>(0,0)}.
\end{equation}
Plugging $t_2 = 0$ into Eq.\eqref{KB4}, we get 
\begin{align}
    \ci \partial_{t_1}G^{>,<}(t_1) &= J_q^2 \int_0^{t_1}dt [G^>(t) G^<(t)]^{q/2 - 1} \nonumber \\
    & \times [G^>(t) - G^<(t)] G^{>,<}(t_1 - t).
\end{align}
We further define, $G^>(t)= h_1(t) G^>(0) = -h_1(t) \ci (1-f)$ and $G^<(t) = h_2(t) G^<(0) = h_2(t) \ci f$, to obtain
\begin{subequations}
\begin{align}
    \partial_t h_1(t) &= - \frac{2}{q}\mathcal{J}_q^2 \int_0^t dt_1 [h_1(t_1) h_2(t_1)]^{\frac{q}{2} - 1}\nonumber \\
    &\times [(1-f)h_1(t_1) + fh_2(t_1)] h_1(t - t_1), \\ 
    \partial_t h_2(t) &= - \frac{2}{q}\mathcal{J}_q^2 \int_0^t dt_1 [h_1(t_1) h_2(t_1)]^{\frac{q}{2} - 1} \nonumber \\
    &\times [(1-f)h_1(t_1) + fh_2(t_1)] h_2(t - t_1),
\end{align}
\end{subequations}
with $\mathcal{J}_q^2 = (q/2)J_q^2 [f(1-f)]^{q/2 - 1} $. Due to the symmetry of the equations and the initial conditions $h_1(0) = h_2(0) = 1$, it is easy to show that $h_1(t) = h_2(t) \equiv h(t)$, where 
\begin{equation}
\partial_t h(t) = - \frac{2}{q}\mathcal{J}_q^2 \int_0^t dt_1 h(t_1)^{q-1} h(t - t_1) \label{eq:SimplifiedKB}
\end{equation}
For the SYK$_2$+SYK$_4$ model, the KB equations reduces to 
\begin{equation}
    \partial_t h(t) = -  \int_0^t dt_1 [J_4^2f(1-f)h(t_1)^3 + J_2^2 h(t_1)] h(t - t_1) \label{eq:KB_simpli}
\end{equation}}
\redtext{The above equations can be solved for $h(t)$, with the boundary condition $h(0)=1$, to obtain all the Green's functions and the time evolution of initial density inhomogeneity,
\begin{subequations}\label{eqs:KB_simpli}
\begin{align}
    G^>(t_1, t_2) &= -\ci (1-f) h(t_1-t_2), \\
    G^<(t_1, t_2) &= \ci f h(t_1-t_2),\\
    G_e^>(t_1, t_2) &= -\ci (1-f) h(t_1 -t_2) - \ci f h(t_1) h(t_2,)\\
    G_e^<(t_1, t_2) &= \ci f [h(t_1 - t_2) - h(t_1) h(t_2)],\\
    n_e(t) &= f[1-h(t)^2].
\end{align}
\end{subequations}
We have verified (not shown) that the numerical solution Eq.\eqref{eq:SimplifiedKB} reproduces the results obtained from the full KB equations [Eqs.\eqref{KBEqns1},\eqref{eq:InitialCondition_Ge}], e.g, for $q = 4$, such as in Fig.\ref{fig:SYKDensityDependence}.}

\subsubsection{\redtext{Exact solution of the pure-state Kadanoff-Baym equations for $\mathrm{SYK}_2$ model}}\label{sec:KBAnalytic_SYK2}
\redtext{The simplified KB Eq.\eqref{eq:SimplifiedKB} for the $\mathrm{SYK}_2$ model,
\begin{equation}
    \partial_th(t) = - J_2^2 \int_0^t dt h(t_1)h(t-t_1),
\end{equation}
is  exactly solvable through Laplace transform $\tilde{h}(s)$ of $h(t)$.
The Laplace transform of the above equation is $s\tilde{h}(s) - 1 = - \tilde{h}(s)^2$, giving us $\tilde{h}(s) = [-s + \sqrt{s^2+4}]/2$. Applying inverse Laplace transform yields, 
\begin{equation}
    h(t) = \frac{J_1(2t)}{t}.
\end{equation}
This expression, along with Eqs.\eqref{eqs:KB_simpli}, matches exactly with the single-particle RMT results derived below.
}
\subsubsection{\redtext{Late-time solution of the pure-state Kadanoff-Baym equations for $\mathrm{SYK}_{q\geq 4}$ model}}\label{sec:KBAnalytic_SYK4}

\redtext{The asymptotic late-time solution of the simplified KB Eq.\eqref{eq:SimplifiedKB} for $q\geq 4$ is obtained by substituting $h(t)=e^{-\gamma_q t}$ to get
\begin{align}
    \gamma_q^2=\frac{2\mathcal{J}_q^2}{q(q-2)}(1-e^{-(q-2)\gamma_q t})
\end{align}
In the limit $\gamma_q t\gg 1$, the above leads to $\gamma_q=\sqrt{2/q(q-2)}\mathcal{J}_q$, and
\begin{align}
    h(t)\simeq e^{-\sqrt{2/q(q-2)}\mathcal{J}_q t}. \label{eq:Late-time_h}
\end{align}
In this late time regime, the Green's functions $G^{>,<}(t_1,t_2)$, $G_e^{>,<}(t_1,t_2)$ and $n_e(t)$ can be obtained by substituting the above $h(t)$ into Eqs.\eqref{eqs:KB_simpli}.}

\redtext{We note that the above late-time exponential ansatz $h(t)=e^{-\gamma_q t}$ does not work for $q=2$ in Eq.\eqref{eq:SimplifiedKB} and for $\mathrm{SYK}_2+\mathrm{SYK}_4$ model in Eq.\eqref{eq:KB_simpli}. 
}

\subsubsection{\redtext{Solutions of pure-state Kadanoff-Baym equations for $\mathrm{SYK}_q$ model in the large-$q$ limit}}\label{sec:KBAnalytic_Largeq}

In this subsection, we consider the SYK$_q$ model defined in Eq. \eqref{SYKqHam}. The large-$N$ KB equations [Eqs.\eqref{KBEqns1}, Appendix \ref{SK_actionKBEqns}] of this model are analytically solvable in the limit of large $q$. The self-energy of this model is given by 
\begin{equation}
    \Sigma^{>,<}(t_1,t_2) = J_q^2 G^{>, <}(t_1, t_2)^{q/2} G^{<, >}(t_2, t_1)^{q/2 - 1}\label{selfenergy}. 
\end{equation}
We will follow the large-$q$ analysis, as is commonly done in the literature \cite{Eberlein2017, SYK2016, Louw_manySYK}.  At $q=\infty$, the theory becomes a free theory i.e., $\Sigma = 0$. We write $1/q$ expansion of all the Green's functions in our large-$N$ theory as 
\begin{equation}
    G(t_1, t_2) = G_0(t_1, t_2) \left(1 + \frac{g(t_1,t_2)}{q} + O\left(\frac{1}{q^2}\right)\right)
\end{equation}
where $G_0(t_1, t_2)$ is the Green's function of the free theory. See Eqs.\eqref{largeqG} and Eqs.\eqref{eq:largeq_g} of Appendix \ref{App:largeqSYK} for details. The Green's functions of complex fermions satisfy the conjugacy property $G^{>,<}(t_1,t_2)^* = - G^{>,<}(t_2, t_1)$, and consequently the $g$'s satisfy $g(t_1, t_2) = g(t_2, t_1)^*$.

Substituting the Green's function expansions from Eqs.\eqref{largeqG} into the self-energy in Eq.\eqref{selfenergy} and requiring that the first non-zero term is at order $(1/q)$, we get 
\begin{subequations}
\begin{align}
    \Sigma^>(t_1, t_2) &= -\frac{\ci}{q}2\mathcal{J}_q^2 (1-f) e^{g_+(t_1, t_2)} \\
    \Sigma^<(t_1, t_2) &= \frac{\ci}{q}2\mathcal{J}_q^2 f e^{g_+(t_2, t_1)}
\end{align}
\end{subequations}
where we define the symmetrized function $g_+(t_1, t_2) \equiv [g^>(t_1, t_2)+g^<(t_2,t_1)]/2$ and $J_q$ is scaled with $q$ such that $\mathcal{J}_q^2 = J_q^2 (q/2) [f(1-f)]^{q/2 -1}$ is independent of $q$. The KB equations at $\mathcal{O}(1/q)$ lead to first-order differential equations [Eqs.\eqref{eq:largeq_KBg}] for $g_e^{>,<}(t_1, t_2)$ in $t_1, t_2$, as listed in Appendix \ref{App:largeqSYK}. Taking appropriate time derivatives and using the definition of $g_+(t_1, t_2)$ give 
\begin{equation}
    \partial_{t_2}\partial_{t_1}g_+(t_1, t_2) = 2\mathcal{J}_q^2 e^{g_+(t_1,t_2)}. \label{LiouvilleEquation}
\end{equation}
This is a Liouville equation and the general solution to this equation, satisfying the conjugacy property of $g$ is given by \cite{TSUTSUMI1980}
\begin{equation}
    e^{g_+(t_1, t_2)} = -  \frac{\dot{u}(t_1)\dot{u}^*(t_2)}{\mathcal{J}_q^2 [u(t_1) - u^*(t_2)]^2} \label{GenSolution}.
\end{equation}
Substituting this general solution into the KB equations at $1/q$ order gives us 
\begin{equation}
    \frac{\ddot{u}(t_1)}{\dot{u}(t_1)} = \frac{\dot{u}(t_1)}{u(t_1) - u^*(0)} + \frac{\dot{u}(t_1)-\dot{u}^*(t_1)}{u(t_1) - u^*(t_1)} - \frac{\dot{u}^*(t_1)}{u^*(t_1) - u(0)} \label{u2eq}
\end{equation}
We observe that $u(t) = \ci e^{\sigma t}$ is a solution to the above equation. An important property of Eq.\eqref{GenSolution} is its invariance under $\mathrm{SL}(2, \mathbb{R})$ transformation $u(t) \rightarrow \frac{au(t)+b}{c u(t) + d}$ where $a,b,c,d \in \mathbb{R}, ad-bc = 1$. Exploiting this property, we make the ansatz 
\begin{equation}
    u(t) = \frac{a \ci e^{\sigma t} + b}{c \ci e^{\sigma t} + d}.
\end{equation}
Substituting this ansatz in the Liouville equation gives us, $\sigma = 2 \mathcal{J}_q$ and
\begin{equation}
    g_+(t_1, t_2) =  -2\log[\cosh (\mathcal{J}_q(t_1-t_2))].
\end{equation}
Using this result, we integrate the first-order differential equations of $g_{e}(t_1, t_2)$ [Eqs.\eqref{eq:largeq_KBg}] to obtain the following solutions to the leading order in $1/q$
\begin{subequations}\label{eqs:large_qsol}
\begin{align}
    G_e^<(t_1, t_2) &= \frac{\ci}{q}[ 2f \log(\cosh \mathcal{J}_q t_1 \cosh \mathcal{J}_q t_2) \nonumber\\ 
    & \hfill - 2f \log(\cosh \mathcal{J}_q (t_1 -t_2)) ] \label{eq:Largeq_Ge<} \\
    G_e^>(t_1, t_2) &= -\ci + \frac{\ci}{q}[ 2f \log(\cosh \mathcal{J}_q t_1 \cosh \mathcal{J}_q t_2)  \nonumber \\
    &+ 2(1-f) \log(\cosh \mathcal{J}_q (t_1 -t_2)) ] \label{eq:Largeq_Ge>}\\
    G^>(t_1, t_2) &= -\ci(1-f)[1 - \frac{2}{q} \log \cosh(\mathcal{J}_q(t_1-t_2))] \\ 
    G^<(t_1, t_2) &= \ci f [ 1 - \frac{2}{q} \log \cosh(\mathcal{J}_q(t_1-t_2)) ] \\
    n_e(t) &= \frac{4f}{q} \log \cosh \mathcal{J}_q t \label{Largeqnc}.
\end{align}
\end{subequations}
We observe that the large-$N$ Green's functions $G(t_1, t_2)$ reach a steady state instantaneously, i.e., they become time-translationally invariant, $G(t_1, t_2) = G(t_1-t_2)$ for all $t_1,t_2$. Applying the Kubo–Martin–Schwinger (KMS) condition, $G^<(t+\ci \beta)/G^>(t) = - e^{\beta \mu} $ with chemical potential $\mu$, we find that $\beta=0$, i.e., temperature is infinity, as expected from ETH for the pure Fock state, that corresponds to the energy expectation $\langle n|\mathcal{H}_q|n\rangle=0$ in the large-$N$ SYK model. Thus, the pure Fock states lies at the middle of the many-body energy spectrum of the SYK model \cite{ChowdhurySYKReview}. The infinite temperature along with the KMS condition, imply that the chemical potential also tends to infinite, such that $e^{\beta \mu}\to f/(1-f)$, a finite value determined by the filling. However, the local Green’s functions $G_{e/f}(t_1, t_2) $ do not thermalize instantaneously, as can be seen from Eqs.\eqref{eq:Largeq_Ge<},\eqref{eq:Largeq_Ge>}. This is in contrast to the quench to a large-$q$ SYK Hamiltonian from initial homogeneous states that are either thermal equilibrium states of some combination of static SYK Hamiltonians, or thermal equilibrium states evolving under combinations of time-dependent SYK Hamiltonians \cite{Eberlein2017, Louw_manySYK}. In case of homogeneous mixed initial states, local Green's functions are identical to the large-$N$ Green's functions, and instantaneous thermalization was found for all the Green's functions up to $\mathcal{O}(1/q)$ in the large-$q$ limit. 

The large-$q$ result for $n_e(t)$ in Eq.\eqref{Largeqnc} also provides an estimate of the filling dependence of the thermalization rate for the density inhomogeneity for the pure state for finite $q$. The thermalization time scale can be estimated from the condition $n_e(t)\simeq f$, leading to $\mathcal{J}_qt\sim \mathcal{O}(1)$. This implies that the thermalization rate is approximately given by $\mathcal{J}_q \propto [f(1-f)]^{q/4-1/2}$. \redtext{The time scale obtained from large-$q$ analysis is the same as the one obtained from the late-time asymptotic solution of the KB equations in Sec.\ref{sec:KBAnalytic_SYK4}. Moreover, the large-$q$ result above can be derived by taking the large-$q$ limit of Eqs.\eqref{eq:SimplifiedKB}. Writing $h(t) = 1 + h_1(t)/q + \dots$
yields (upto $1/q$)
\begin{equation}
    \partial_th_1(t) = - 2 \mathcal{J}_q^2 \int_0^t dt_1 e^{h_1(t_1)},
\end{equation}
with $h_1(0) = 0$, whose solution is given by $h_1(t) = -2 \log \cosh \mathcal{J}_qt$. This solution, along with Eqs.\eqref{eqs:KB_simpli} reproduces the large-$q$ results Eqs.\eqref{eqs:large_qsol} above. 
}
\redtext{In Sec.\ref{sec:Imbalance}, we compare our analytical results  with the numerically obtained time-evolution of density inhomogeneity.}

\subsection{Single-particle random matrix theory for pure state evolution in SYK$_2$ model}\label{sec:RMT_singleparticle}
The hopping amplitudes $J_{ij}$ in SYK$_2$ are drawn from the Gaussian Unitary Ensemble (GUE). In the GUE, the eigenvalues $\{\epsilon_\alpha\}$ and eigenfunctions $\{\psi_\alpha(i)\}$ are independently distributed, with the eigenvalues following Wigner's semi-circular distribution and the eigenvectors uniformly distributed according to the Haar measure on the unitary group \cite{mehta1991random, Livan2018}. Using this property, along with Eqs.\eqref{nc} and \eqref{cdagtct} of Appendix \ref{ED_SYK2}, we obtain the disorder-averaged density of the initially empty sites as
\begin{align}
    n_e(t) &= \frac{1}{(1-f)N} \sum_{\substack{i \notin I,j\\\alpha\beta}}n_{j}
    \overline{\psi_\alpha(i) \psi^*_\beta(i) \psi^*_\alpha(j) \psi_\beta(j)}  \nonumber\\
    &\times \overline{e^{\ci(\epsilon_\beta - \epsilon_\alpha) t}}.
\end{align}
Adapting the notation $U_{i \alpha} = \psi_\alpha(i)$, the first average corresponds to the integral over Haar distribution, which can be computed exactly using Weingarten calculus \cite{collins2002momentscumulantspolynomialrandom, Collins_2006}. For our case, it is given by 
\begin{multline}
    \overline{U_{i \alpha}U_{j \beta} U^\dagger_{\alpha' i'} U^\dagger_{\beta' j'}} = \frac{1}{N^2-1}[\delta_{ii'} \delta_{\alpha \alpha'} \delta_{jj'}\delta_{\beta \beta'} \\+ \delta_{ij'} \delta_{\alpha \beta'} \delta_{ji'}\delta_{\beta \alpha'}- \frac{1}{N}(\delta_{ii'} \delta_{\alpha \beta'} \delta_{jj'}\delta_{\beta \alpha'}+\delta_{ij'} \delta_{\alpha \alpha'} \delta_{ji'}\delta_{\beta \beta'})]. \label{eq:RMT_U}
\end{multline}
Using this result, we obtain
\begin{equation}
   n_e(t) =  f \frac{N^2}{N^2-1} [1 - \frac{1}{N^2} \sum_{\alpha \beta} \overline{e^{\ci(\epsilon_\beta - \epsilon_\alpha) t}}].
\end{equation}
The sum appearing in the above expression is the 2-point spectral form factor at infinite temperature and is given by
$\mathcal{R}_2(t)=\sum_{\alpha \beta} \overline{e^{\ci(\epsilon_\beta - \epsilon_\alpha) t}} = N^2 r_1(t)^2 - N r_2(t) +N$
with $r_1(t) \equiv J_1(2t)/t$, $J_1(t)$ is the Bessel function of the first kind \cite{Cotler_2017}, with the unit of time set by the standard deviation of the random matrix elements $J_{ij}$, i.e., $J_2$ in our case. In the large-$N$ limit, only the first term survives at the leading order to give,
\begin{equation}
n_e(t) = f \left[1 - \left(\frac{J_1(2t)}{t}\right)^2\right]. \label{RMTnc}
\end{equation}
The above exactly matches our large-$N$ SK field theory results for $n_e(t)$, as shown in Fig.\ref{fig:SYKDensityDependence}(b). Following the procedure outlined above, we obtain the expressions for the Green's functions
\begin{subequations} \label{eq:G_SingleParticleRMT}
\begin{align}
    G_e^<(t_1, t_2) &= \ci f \frac{J_1(2(t_1 - t_2))}{(t_1 - t_2)} - \ci f \frac{J_1(2t_1)}{t_1} \frac{J_1(2t_2)}{t_2}\\ 
    G_e^>(t_1, t_2) &= -\ci (1-f) \frac{J_1(2(t_1 - t_2))}{(t_1 - t_2)} \nonumber \\
    &- \ci f \frac{J_1(2t_1)}{t_1} \frac{J_1(2t_2)}{t_2}\\ 
    G^<(t_1, t_2) &= \ci f \frac{J_1(2(t_1 - t_2))}{(t_1 - t_2)} \\ 
    G^>(t_1, t_2) &= -\ci (1-f) \frac{J_1(2(t_1 - t_2))}{(t_1 - t_2)}. 
\end{align}
\end{subequations}
\redtext{The Green's functions $G^{>,<}(t_1,t_2)$, $G_e^{>,<}(t_1,t_2)$ and $n_e(t)$ obtained above from single-particle RMT exactly match with the analytical solutions of the KB equations [Eqs.\eqref{KBEqns1},\eqref{eq:InitialCondition_Ge}] for SYK$_2$ in Sec.\ref{sec:KBAnalytic_SYK2}.}


\begin{figure*}[ht]
    \centering
    \includegraphics[width=\textwidth]{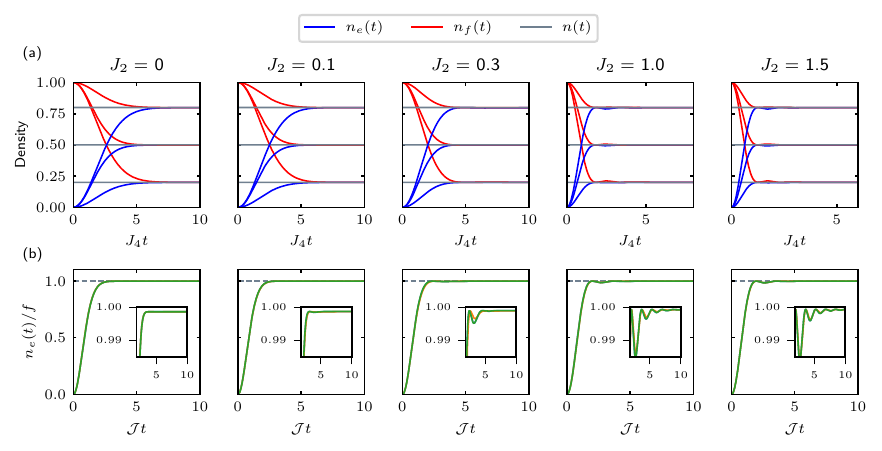}
    \caption{{\bf Thermalization dynamics of density inhomogeneity of the pure state in the SYK$_2$+SYK$_4$ model:} (a) Crossover in the SYK$_2$+SYK$_4$ model from density-dependent monotonic relaxation, like SYK$_4$, to a faster oscillatory relaxation, like SYK$_2$, as $J_2$ increased from left to right. Results for three values of filling, $f=0.2,0.5,0.8$, are shown. (b) Approximate scaling collapse: $n_e(t)/f = f_{24}(\mathcal{J} t)$, where $\mathcal{J} = \sqrt{J_2^2 + J_4^2 f(1-f)}$.}
    \label{fig:SYK2plus4App}
\end{figure*}

\subsection{\redtext{Many-body random matrix theory for pure state evolution under random Hamiltonians}} \label{sec:G_RMT_Thermalization}

\redtext{Using many-body RMT, here we show that instantaneous thermalization of the disorder and site-averaged local Green's function $G(t_1,t_2)$, starting from a generic pure state, is a general property of any random Hamiltonian in the thermodynamic limit. Moreover, we establish a connection between $G(t_1,t_2)$ and quantity similar to the spectral form factor \cite{Brezin1997}.}

We consider an initial pure state $|\Psi(0)\rangle=\sum_n C_n|n\rangle $ in a given particle number sector $fN$. Using the many-body eigen energies and eigen states $\{E_\alpha,|E_\alpha\rangle\}$ of the SYK Hamiltonian $\mathcal{H}$ for given realization of disorder and the expansion $|n\rangle=\sum_\alpha U^\dagger_{n\alpha}|E_\alpha\rangle$, we obtain the time evolved state as
\begin{align}
|\Psi(t)\rangle=\sum_{nm\alpha} C_n U^\dagger_{n\alpha}U_{\alpha m} e^{-\ci E_\alpha t} |m\rangle.
\end{align}
Here $U$ is the many-body unitary matrix that diagonalizes $\mathcal{H}$. Using the above, the Green's function $G_{ii}^>(t_1,t_2)=-\ci \langle \Psi(0)|c_i(t_1)c^\dagger_i(t_2)|\Psi(0)\rangle$, where $c_i(t)=e^{\ci \mathcal{H}t_1}c_ie^{-\ci \mathcal{H}t_1}$, can be written as
\begin{align}
 G_{ii}^>(t_1,t_2)&=-\ci\sum_{\substack{nn'\\\alpha\alpha'\\mm'}}C_n^*C_{n'} U_{\alpha n} U_{\alpha'm'}U^\dagger_{m\alpha}U^\dagger_{n'\alpha'}\nonumber \\
 &\times e^{\ci(E_\alpha t_1-E_{\alpha'}t_2)} \langle m|c_ie^{-\ci \mathcal{H}(t_1-t_2)}c_i^\dagger |m'\rangle.
\end{align}
We define states $c_i^\dagger|m\rangle=(1-m_i)|m(i)\rangle$, $c_i^\dagger|m'\rangle=(1-m_i')|m'(i)\rangle$ that are non-zero when $i$-th site is unoccupied in $|m\rangle, |m'\rangle$. These states correspond to Fock state in the $fN+1$ fermion sector. We write these states, e.g., $|m(i)\rangle=\sum_{\tilde{\alpha}}\tilde{U}^\dagger_{m(i),\tilde{\alpha}}|\tilde{\alpha}\rangle$ in terms of the eigenstates $\{|\tilde{\alpha}\rangle\}$, with energies $\tilde{E}_{\tilde{\alpha}}$, and unitary $\tilde{U}$ of $\mathcal{H}$ in the $fN+1$ fermion sector. As a result, we get
\begin{align}
 G&_{ii}^>(t_1,t_2)\nonumber\\
 &=-\ci\sum_{\substack{nn'\\\alpha\alpha'\tilde{\alpha}\\mm'}}C_n^*C_{n'} U_{\alpha n} U_{\alpha'm'}U^\dagger_{m\alpha}U^\dagger_{n'\alpha'}\tilde{U}_{\tilde{\alpha},m(i)}\tilde{U}^\dagger_{m'(i),\tilde{\alpha}}\nonumber \\
 &\times (1-m_i)(1-m_i')e^{\ci(E_\alpha t_1-E_{\alpha'}t_2)-\ci\tilde{E}_{\tilde{\alpha}}(t_1-t_2)} 
\end{align}
 Writing $t_{1,2}=\mathcal{T}\pm t/2$ in terms of center-of-mass and relative times $\mathcal{T}$ and $t$, we perform disorder averaging over the above Green's function along the same lines as in Sec.\ref{sec:RMT_singleparticle}, where random matrix properties of single-particle unitary and eigen energies were used for disorder averaging.
\begin{align}
 G&_{ii}^>(t_1,t_2)\nonumber\\
 &\approx-\ci\sum_{\substack{nn'\\\alpha\alpha'\tilde{\alpha}\\mm'}} (1-m_i) (1-m_i') C_n^*C_{n'} \overline{U_{\alpha n} U_{\alpha'm'}U^\dagger_{m\alpha}U^\dagger_{n'\alpha'}}\nonumber \\
 &\times \overline{\tilde{U}_{\tilde{\alpha},m(i)}\tilde{U}^\dagger_{m'(i),\tilde{\alpha}}} ~~\overline{e^{\ci(E_\alpha-E_{\alpha'})\mathcal{T}}e^{\ci[(E_\alpha+E_{\alpha'})/2-\tilde{E}_{\tilde{\alpha}}]t}}, 
\end{align}
where we have assumed that unitaries in the $fN$ and $fN+1$ fermion sectors are completely uncorrelated since the Hamiltonian $\mathcal{H}$ is block diagonal in the particle number sectors. Assuming the unitaries to be Haar distributed, we use the same identity of Eq.\eqref{eq:RMT_U} along with $\overline{\tilde{U}_{\tilde{\alpha},m(i)}\tilde{U}^\dagger_{m'(i),\tilde{\alpha}'}}\approx (1/\tilde{\mathcal{N}}_F)\delta_{mm'}\delta_{\tilde{\alpha}\tilde{\alpha}'}$, with $N$ in Eq.\eqref{eq:RMT_U} appropriately replaced by the Fock-space dimensions $\mathcal{N}_F$ and $\tilde{\mathcal{N}}_F$ of the $fN$ and $fN+1$ particle sectors. As a result, we get in the leading order for large $\mathcal{N}_F,\tilde{\mathcal{N}}_F$,
\begin{align}
G&_{ii}^>(\mathcal{T},t)\approx -\frac{\ci}{\mathcal{N}_F^2\tilde{\mathcal{N}}_F}\sum_{\substack{nn'm \\ \alpha\alpha'\tilde{\alpha}}}(1-m_i)C_n^*C_{n'}\left[ (\delta_{nm}\delta_{n'm}\right.\nonumber \\
&\left.+\delta_{\alpha\alpha'}\delta_{nn'})-\frac{1}{\mathcal{N}_F}(\delta_{nn'}+\delta_{\alpha\alpha'}\delta_{nm}\delta_{n'm})\right]\nonumber \\
&\times \overline{e^{\ci(E_\alpha-E_{\alpha'})\mathcal{T}}e^{\ci[(E_\alpha+E_{\alpha'})/2-\tilde{E}_{\tilde{\alpha}}]t}} \nonumber \\
&\simeq -\ci\left[\sum_n(1-n_i)|C_n|^2-\frac{1}{\mathcal{N}_F}\sum_n(1-n_i)\right]\nonumber \\
&\times \frac{1}{\mathcal{N}_F^2\tilde{\mathcal{N}}_F}\sum_{\alpha\alpha'\tilde{\alpha}}\overline{e^{\ci(E_\alpha-E_{\alpha'})\mathcal{T}}e^{\ci\left(\frac{E_\alpha+E_{\alpha'}}{2}-\tilde{E}_{\tilde{\alpha}}\right)t}}\nonumber \\
&-\ci \left[\frac{1}{\mathcal{N}_F}\sum_n(1-n_i)\right] \frac{1}{\mathcal{N}_F\tilde{\mathcal{N}}_F}\sum_{\alpha\tilde{\alpha}}\overline{e^{\ci(E_\alpha-\tilde{E}_{\tilde{\alpha}})t}}
\end{align}
Similar expression can be obtained for $G_{ii}^<(\mathcal{T},t)$. As evident from the above, the local Green's functions depend on the center-of-mass time and do not thermalize instantly. However, the large-$N$ Green's function obtained after averaging over all sites becomes independent of $\mathcal{T}$ and is given by
\begin{align}\label{eq:Ggreater_manybodyRMT}
G^>(t)&=\frac{1}{N}\sum_i G_{ii}^>(\mathcal{T},t)=-\ci  \frac{(1-f)}{\mathcal{N}_F\tilde{\mathcal{N}}_F}\sum_{\alpha\tilde{\alpha}}\overline{e^{\ci(E_\alpha-\tilde{E}_{\tilde{\alpha}})t}}.
\end{align}
It can be verified that the above, and a similar expression for $G^<(t)$, correspond to infinite temperature thermal Green's functions. \redtext{The above expression establishes a connection between site and disorder-averaged local Green's function in the dynamics of a generic pure state under random-Hamiltonian evolution with the quantity $(1/\mathcal{N}_F\tilde{\mathcal{N}}_F)\sum_{\alpha\tilde{\alpha}}\overline{\exp[\mathrm{i}(E_\alpha-\tilde{E}_{\tilde{\alpha}})t]}$. This quantity is similar to spectral form factor (SFF) \cite{Brezin1997}, namely it captures correlations between energy eigenvalues, albeit, here, between many-body energy eigenvalues in two different particle number sectors.} Our SK field theory allows us to exactly calculate $G^{>,<}$ as well as non-equilibrium local and non-local two and multi-point correlators, for any pure state in the large-$N$ limit. 

\subsection{Density dependence of thermalization of initial inhomogeneity}\label{sec:Imbalance}

The time evolution of the local densities satisfies $f n_f(t) + (1-f) n_e(t)=f$, as can be deduced from Eq. \ref{KB3} of Appendix \ref{SK_actionKBEqns}. This implies
\begin{equation}
\frac{n_e(t)}{f} = \frac{1-n_f(t)}{1-f}. 
\end{equation}
This relation, which connects the dynamics of initially unfilled sites and filled sites is used below to collapse the data for $n_e(t)/f$ at different fillings. 

In the SYK$_2$ model, $n_e(t)/f$ for various initial fillings collapses onto a single curve, as shown in Fig.\ref{fig:SYKDensityDependence}(b). This collapse indicates that the dynamics is independent of the filling and match precisely with the result from \redtext{the analytical solution of the KB equations [Sec.\ref{sec:KBAnalytic_SYK2}] and RMT [Sec.\ref{sec:RMT_singleparticle}]}, as given by Eq.\eqref{RMTnc}. \redtext{Performing a taylor expansion about $t = 0$, we find the early-time behavior as,
\begin{equation}
    n_e(t) \sim f (J_2t)^2.
\end{equation}}
Using the asymptotic form of the Bessel function at late times, we find \redtext{the late-time behavior as,}
\begin{equation}
n_e(t) \sim f \left(1 - \frac{1}{(J_2t)^3}\cos^2(2J_2t - 3 \pi/4) \right), \label{eq:LateTimeImbalance_SYK2}
\end{equation}
which suggests that the oscillation amplitudes decay as $(J_2t)^{-3}$ and that the oscillation period is $1/(2J_2)$. 

For the interacting SYK$_{q\geq 4}$, \redtext{from the late-time [Sec.\ref{sec:KBAnalytic_SYK4}] and large-$q$ [Sec.\ref{sec:KBAnalytic_Largeq}] solutions of the KB equations, we found an emergent density-dependent inverse time scale $\propto \mathcal{J}_q\equiv \mathcal{J}= J_q \sqrt{(q/2)} [f(1-f)]^{q/4-1/2}$}. By scaling time with this emergent time scale for $q=4$, $n_e(t)/f$ for various initial fillings collapses onto a single curve in the SYK$_4$ model, as shown in Fig.\ref{fig:SYKDensityDependence}(d). The large-$q$ result in Eq.\eqref{Largeqnc} with $q=4$ shows a good agreement with the large-$N$ predictor-corrector solution of the KB equations in the early time regime $\mathcal{J}_4 t \lesssim O(1)$. \redtext{Similarly, we observe the collapse for all finite $q \geq 4$ when the time is scaled by $\mathcal{J}_q$, as shown in Fig.\ref{fig:SYKqScaling} in Appendix \ref{App:largeqSYK}}. 

\redtext{To analyze the early-time behavior of $n_e(t)$ for SYK$_4$, we perform a Taylor expansion of $h(t)$ in Eq.\eqref{eq:SimplifiedKB} for $q=4$ about $t=0$ Eq.\eqref{eq:SimplifiedKB} and obtain
\begin{equation}
    n_e(t)=f[1-h(t)^2] \sim f \frac{(\mathcal{J}_4t)^2}{2}. 
\end{equation}
The late-time behavior is obtained by using Eq.\eqref{eq:Late-time_h} to get
\begin{equation}\label{eq:ne_latetime_SYK4}
    n_e(t) \sim f(1 - e^{-\mathcal{J}_4t}) 
\end{equation}
where $\mathcal{J}_4 = \sqrt{2f(1-f)}J_4$. This exponential decay, green curve in Fig.\ref{fig:SYKDensityDependence}(d), also agrees with the numerical data in the late-time regime.}

\redtext{The late-time power-law decay of imbalance [Eq.\eqref{eq:LateTimeImbalance_SYK2}] in the $\mathrm{SYK}_2$ model and exponential decay [Eq.\eqref{eq:ne_latetime_SYK4}] in the $\mathrm{SYK}_{q\geq 4}$ is one of the major differences between the non-equilibrium dynamics of pure state in the non-interacting and interacting cases.}

\begin{figure}[ht]
    \centering
    \includegraphics[width=\linewidth]{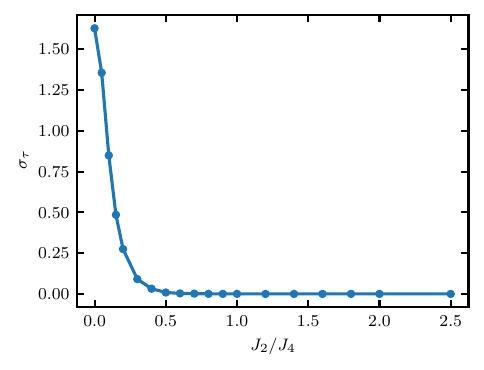}
    \caption{{\bf Density dependence of the thermalization dynamics in the SYK$_2$+SYK$_4$ model:} Variance of the thermalization time scales over fillings in the SYK$_2$+SYK$_4$ model as a function of the relative coupling strength $J_2/J_4$.}
    \label{fig:SYK24sigma}
\end{figure}

\subsubsection{Crossover in SYK$_2$+SYK$_4$ model}
In this subsection, we study the dynamics of pure Fock state in the SYK$_2$+SYK$_4$ model. The self-energy of this model is sum of self-energies of the SYK$_2$ and SYK$_4$ models 
\begin{align}
    \Sigma^{>,<}(t_1, t_2) &= J_2^2 G^{>,<}(t_1, t_2) \nonumber \\
    &+ J_4^2 G^{>,<}(t_1, t_2)^2 G^{<,>}(t_2, t_1),
\end{align}
with the rest of the KB equations [Eqs.\eqref{KBEqns1}, Appendix \ref{SK_actionKBEqns}] being same as the SYK$_q$ model. In the SYK$_2$+SYK$_4$ model, we expect the the thermalization to be density independent for $J_2\gg J_4$ and density dependent for $J_4\gg J_2$. The thermalization dynamics of the density inhomogeneity in the initial pure Fock state is shown in Figs.\ref{fig:SYK2plus4App}(a) for different $J_2/J_4$ values at three different fillings. In case of density dependent relaxation of the imbalance, the thermalization time varies with filling $f$. We numerically calculate the variance of the characteristic time scale over nine different fillings between 0 and 1 by estimating the time when $n_e$ and $n_f$ first reaches the initial filling $f$. Fig.\ref{fig:SYK24sigma} shows the variance as a function of relative coupling strength $J_2/J_4$, which exhibits exponential decay without any non-analyticity. This indicates a crossover, as opposed to a transition, from density-independent oscillatory decay of the initial inhomogeneity in the SYK$_2$ limit of $J_2\gg J_4$ to density-dependent monotonic relaxation for $J_4 \gg J_2$ in the SYK$_4$ limit.

Next, we investigate the scaling collapse in this model. To estimate the thermalization time scale on density and $J_2/J_4$ ratio, we perform a short-time expansion of the KB equations. We carry out a single step of predictor-corrector algorithm, i.e., one $\Delta t$ step along $t_1$, $t_2$ and the diagonal. We use trapezoidal rule to calculate the integrals in the KB equations to obtain 
\begin{subequations}
\begin{align}
    G_e^<(\Delta t, 0) = 0,& \  G_e^>(\Delta t, 0) = -\ci + \frac{\ci}{2} (\mathcal{J} \Delta t)^2 \\
    G_e^<(0, \Delta t) = 0,& \  G_e^>(0, \Delta t) = -\ci + \frac{\ci}{2} (\mathcal{J} \Delta t)^2 \\
    G_e^<(\Delta t, \Delta t) = \ci f (\mathcal{J} \Delta t)^2 ,& \ G_e^>(\Delta t, \Delta t) = -\ci + \ci f (\mathcal{J} \Delta t)^2   \\
    \mathcal{J} &= \sqrt{J_2^2 + J_4^2 f(1-f)}.
\end{align}
\end{subequations}
By scaling the dynamics with the inverse time scale $\mathcal{J}$, we observe an approximate scaling collapse, as shown in Fig.\ref{fig:SYK2plus4App}(b). 

\begin{figure}[ht]
    \centering
    \includegraphics[width=\linewidth]{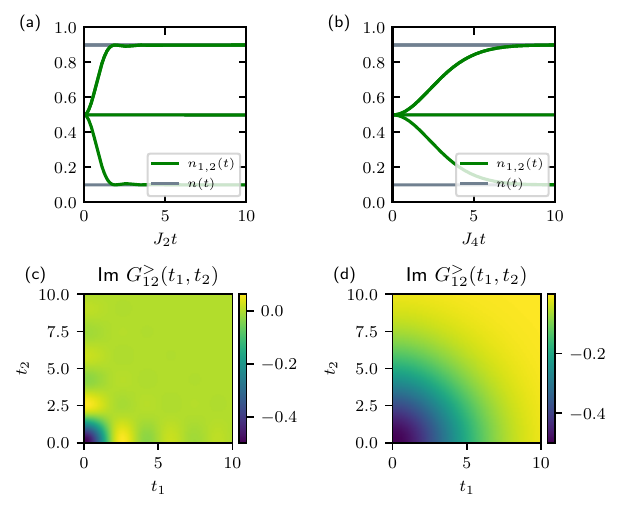}
    \caption{{\bf Thermalization of a Bell pair state in the SYK$_2$ and SYK$_4$ models:} Relaxation of local densities of Bell pair sites $n_{1,2}(t)$ in the initial Bell pair state [Eq.\eqref{eq:BellPairState}] for various fillings, $f=0.1, 0.5, 0.9$ for (a) SYK$_2$ and (b) SYK$_4$ models. Non-local correlations of Bell pair state [Eq.\eqref{eq:SYKBellPairNonLocalCorrel}] in (c) SYK$_2$ and (d) SYK$_4$ models at half-filling ($f=1/2$).}
    \label{fig:SYKBellPair}
\end{figure}

\subsection{Time evolution of an initial entangled Bell pair state in SYK$_2$ and SYK$_4$ models and non-local correlations}
Here we discuss the non-equilibrium dynamics of a simple initial entangled state of Eq.\eqref{eq:BellPairState}, where two entangled sites $i=1,2$ (arbitrary) in a Bell pair state $(1/\sqrt{2})(|10\rangle+|01\rangle)$ is immersed in a product state of other fermions. Under the SYK dynamics the non-local quantum correlation due to this entanglement is lost as the system thermalizes.  
The time evolution of local densities of the Bell pair initial state can be obtained from Eqs.\eqref{eq:BellPair_localG}, \eqref{eq:BellPair_localGi} to give $
n_1(t) = n_2(t) = n_e(t) - \frac{1}{2}G_e^>(t,0)G_e^>(0,t)$ and $n_{i \neq 1,2} = n_e(t) - n_i G_e^>(t,0)G_e^>(0,t)$ for the other sites. Recall that for a pure Fock product state, $n_f(t) = n_e(t) - G_e^>(t,0)G_e^>(0,t) $ which gives us 
\begin{subequations}\label{eq:SYKBellPairDensities}
    \begin{align}
        n_1(t) = n_2(t) &= \frac{1}{2}(n_e(t) + n_f(t)) \\
        n_{i \neq 1,2}(t) &= (1-n_i) n_e(t) + n_i n_f(t).
    \end{align}
\end{subequations}

We plot the time evolution of the local densities for the Bell pair sites in Fig.\ref{fig:SYKBellPair}(a),(b) for three different fillings. The dynamics of the local densities at other sites are same as those in Fig.\ref{fig:SYKDensityDependence}.

Next, we obtain the non-local correlations $G_{12}, G_{21}$ of Eq.\eqref{eq:BellPair_Corrlation} in real time by setting $z_1 \rightarrow t_{1} +(-), z_2 \rightarrow t_2 -(+)$ to get the lesser $G^<$ (greater $G^>$) functions
\begin{equation}\label{eq:SYKBellPairNonLocalCorrel}
    G_{12}^{>,<}(t_1, t_2) = G_{21}^{>,<}(t_1, t_2) = \frac{\ci}{2} G_e^>(t_1, 0) G_e^>(0, t_2).
\end{equation}
In Fig.\ref{fig:SYKBellPair}, we plot the time evolution of non-local correlations $G_{12}^>(t_1,t_2)$ for the Bell pair initial state at half-filling in the SYK$_2$ and SYK$_4$ models. The non-local correlations are obtain straightforwardly from the numerical solutions of $G_e$ shown, e.g., in Fig.\ref{fig:SYKGreensFn}. The non-local correlations which encode the initial entanglement of the Bell pair, decays in a oscillatory fashion for SYK$_2$ model in Fig.\ref{fig:SYKBellPair}(c). In the interacting SYK$_4$ model, the correlations decay monotonically, as shown in Fig.\ref{fig:SYKBellPair}(d).


\section{Conclusions} \label{Conclusion}
In this paper, we developed a new Schwinger-Keldysh (SK) formalism to study the non-equilibrium dynamics of a general pure state with a fixed number of fermions. Our method is exactly solvable in the limit of large-$N$ for the SYK models. Through our large-$N$ SK theory for SYK model, we unravel an intriguing universality in the non-equilibrium dynamics of pure states in the SYK model, where local and non-local correlations of almost all pure states in a fixed particle number sector are completely determined by a single universal Green's function. We show that irrespective of the initial state and the type of the SYK model, i.e., interacting or non-interacting, the site- and disorder-averaged large-$N$ Green's function thermalizes instantaneously to an infinite-temperature Green's function, whereas local and non-local Green's functions have finite thermalization rate due to inhomogeneity in the initial state. \redtext{We support our large-$N$ SK field theory numerical results through exact analytical solutions of the Kadanoff-Baym equations for the pure-state evolutions in the $\mathrm{SYK}_2$ model and large-$q$ $\mathrm{SYK}_q$ model, as well as through asymptotic late-time solution for general $q\geq 4$. We provide further understanding of our results based on both many-body and single-particle random-matrix theory (RMT).} Our results revealed interesting contrasts, quite distinct from that in the low-temperature FL and NFL regimes \cite{Eberlein2017,jaramillo2024thermalizationclosedsachdevyekitaevthermodynamic,Louw_manySYK}, between thermalization dynamics in interacting and non-interacting SYK models. For instance, we find that large-$q$ SYK is not a perfect thermalizer for pure states, i.e., it does not instantaneously thermalize local Green’s functions.

\redtext{It is interesting to ask how finite-$N$ corrections modify the rate of thermalization of the (large-$N$) site-averaged Green's function $G^{>,<}(t_1,t_2)$ from infinity (instantaneous) to a finite value and how the rate depends on $N$ while approaching the thermodynamic limit $N\to \infty$. These are beyond the scope of the large-$N$ SK field-theory methods employed in this work. A few works \cite{Bandyopadhyay2023,DelayedThermMassDefSYK,Dieplinger2023} have numerically studied the thermalization in SYK models at finite $N$, but have not looked into $G^{>,<}(t_1,t_2)$. As a result, we cannot compare directly with the these works \cite{Bandyopadhyay2023,DelayedThermMassDefSYK,Dieplinger2023}. However, Ref.\cite{DelayedThermMassDefSYK} studies chaos and the RMT behavior in a mass-deformed SYK model ($\mathrm{SYK}_2+\mathrm{SYK}_4$ model) through the spectral form factor (SFF). The authors find (Fig.2(d), Ref.\cite{DelayedThermMassDefSYK}) that the Thouless time, i.e., the time scale after which RMT behavior emerges in the SFF, decreases rapidly with $N$ to a very small value, even for moderate $N$ like $N=17$ ($N=34$ Majorana sites). Given the connection between the site and disordered averaged large-$N$ Green's function and SFF, discussed in Sec.\ref{sec:G_RMT_Thermalization}, the above observation is consistent with instantaneous thermalization of the large-$N$ Green's function we find for $N\to \infty$. Moreover, Ref.\cite{Bandyopadhyay2023} numerically found an universality in the equilibration dynamics of disorder-averaged few-body observables for the evolution under the $\mathrm{SYK}_4$ Hamiltonian starting from various distinct initial entangled pure states. The universality in the large-$N$ dynamics found in our work for any generic initial state provides natural rationalization of the above observation. Ref.\cite{Dieplinger2023} studied thermalization through local density-density autocorrelation in the mass-deformed SYK model and found a finite $\mathcal{O}(J)$ relaxation time scale for the local density for the infinite-temperature mixed state. This is again consistent with the $\mathcal{O}(J)$ relaxation time we find for local Green's functions and initial density inhomogeneity of the infinite-temperature pure states in our work. We also note that, unlike our work, all the works  \cite{Bandyopadhyay2023,DelayedThermMassDefSYK,Dieplinger2023} mentioned above only studied the the half-filled SYK models and did not explore the filling dependence of the thermalization.}


The pure Fock product state studied here are same as the infinite-temperature KM states \cite{kourkoulou2017purestatessykmodel,ZhangPureStateSYKdynamics} of the Majoran fermions, if we map the complex fermions $\{c_i,c_i^\dagger\}$ with $i=1,\cdots,N$ to the Majoran fermions $\chi_j$ with $j=1,\cdots, 2N$, through $c_i=(\chi_{2i-1}+\ci \chi_{2i})/2$ and its hermitian conjugate. However, the dynamics in the Majorana model does not conserve the fermion number, unlike the complex SYK models considered here. Nonetheless, the dynamics for the half-filled complex SYK model is identical to that of the Majorana SYK model in the large-$N$ limit, once the complex fermion correlations are translated to the Majorana fermion correlations using the fermion to Majorana mapping. The complex SYK allows us to study the dynamics away from half filling and reveal density-dependent thermalization. It will be worthwhile to extend our general SK formalism and the large-$N$ theory for SYK models to study low-energy pure states generated by an imaginary-time ($\tau$) evolution, e.g., generate the initial states $\sim e^{-\tau \mathcal{H}}|n\rangle$ through the SYK Hamiltonians $\mathcal{H}$, as in the case of KM states \cite{kourkoulou2017purestatessykmodel,ZhangPureStateSYKdynamics}. Initial pure states like $\sim e^{-\tau \mathcal{H}}|n\rangle$ are correlated with SYK Hamiltonian and are not described by the universal dynamics discussed in this work.
Our approach can be generalized to study dynamics of R\'enyi entanglement entropy using the methods of Refs.\onlinecite{Haldar2020,Bera2024}, which we leave for a future work. More broadly, our method opens several new directions in quantum dynamics, such as investigating quantum chaos in pure states \cite{SYKPureStateChaos}, dynamics of non-stabilizerness or quantum magic \cite{bera2025nonstabilizernesssachdevyekitaevmodel}, in the large-$N$ limit, which have previously been studied only for small and finite system sizes.

\textbf{\emph{Data and code availability}.---} Data and code are available on Zenodo \cite{Zenodo}.

\begin{acknowledgments}
We thank Thomas Scaffidi for valuable discussions during the final stages of this work, particularly for suggesting the RMT calculation. RP acknowledges support from the Kishore Vaigyanik Protsahan Yojana, Department of Science and Technology, Government of India. SB acknowledges support from CRG, SERB (ANRF), DST, India (File No. CRG/2022/001062) and STARS, MoE, Govt. of India (File. No. MoE-STARS/STARS-2/2023-0716). AH acknowledges support from DST India via the SERB (ANRF) grant SRG/2023/000118.
\end{acknowledgments}

\appendix
\section{Disorder-averaged large-$N$ Kadanoff-Baym equations for pure-state evolution in SYK model} \label{SK_actionKBEqns}
\subsection{Initial pure product state}
In this appendix, we derive the large-$N$ Schwinger-Keldysh (SK) $G-\Sigma$ action of the SYK$_q$ model in Eq.\eqref{SYKqHam} of Sec.\ref{models} for time evolution from an initial pure product (Fock) state $|n\rangle$. Using, the general SK formalism of Sec.\ref{SKPureStateFormalism}, the SK action [Eq.\eqref{eq:SKAction_FockState}] for this model is given by 
\begin{multline}
    S = \int_{\mathcal{C}} dz \sum_i  \bar{c}_i(z) \ci\partial_z  c_i(z) \\ - \int_{\mathcal{C}} dz \sum_{\substack{i_1, \dots, i_{q/2} \\ j_1, \dots j_{q/2}}} J_{i_1 \cdots,i_{q/2}; j_1 \cdots j_{q/2}} \bar{c}_{i_{q/2}} \cdots \bar{c}_{i_{1}} c_{j_1} \cdots c_{j_{q/2}}\\
     -\ci \int_{\mathcal{C}} dz \sum_{i \in I} [\bar{c}_i(z)\delta_{\mathcal{C}}(z,0+)\eta_i - \bar{\eta}_i \delta_{\mathcal{C}}(z,0-)c_i(z) ],
\end{multline}
where the last term incorporates the initial pure state in this formalism. We perform disorder-averaging of the generating function $Z$ over the Gaussian random couplings $J_{i_1 \cdots i_{q/2}; j_1 \cdots j_{q/2}}$ to get the following action, 
\begin{multline}
    S = \int_{\mathcal{C}} dz_1 dz_2 \\ \left[\sum_i  \bar{c}_i(z) [\ci\partial_{z_1}\delta_{\mathcal{C}}(z_1-z_2) - \Sigma(z_1,z_2)  ]c_i(z) \right.\\ \left.-\ci N\Sigma(z_1,z_2)G(z_2,z_1)+ \ci \frac{J_q^2}{q}N G(z_1,z_2)^{q/2}G(z_2,z_1)^{q/2}\right] \\
     -\ci \int_{\mathcal{C}} dz \sum_{i \in I} [\bar{c}_i(z)\delta_{\mathcal{C}}(z,0+)\eta_i - \bar{\eta}_i \delta_{\mathcal{C}}(z,0-)c_i(z) ].
\end{multline}
In the above, we have introduced the large-$N$ field $G(z_1,z_2)$ and its conjugate field $\Sigma(z_1,z_2)$ as a Lagrange multiplier, i.e.,
\begin{equation}
    G(z_1,z_2) = -\frac{\ci}{N}\sum_{i}c_{i}(z_1)\bar{c}_{i}(z_2) 
\end{equation}
\begin{multline}
    1 = \int \mathcal{D}(G, \Sigma) \exp\left[\int_{\mathcal{C}}dz_1 dz_2 \Sigma(z_2, z_1)\times \right. \\ \left. \left(N G(z_1, z_2) + \ci \sum_i c_i(z_1) \bar{c_i}(z_2)\right)\right].
\end{multline}
After integrating out the fermions, we obtain the disorder-averaged generating function in the $G-\Sigma$ representation $\overline{Z} = \int\mathcal{D}(G, \Sigma) e^{\ci (S_{c}+S_{G})}
$, where
\begin{subequations} \label{eq:SKAction_SYKFock}
\begin{align}
S_{G}[G,\Sigma] &= \ci N\int_{\mathcal{C}} dz_1 dz_2 \left[\frac{J_q^{2}}{q}G(z_1,z_2)^{q/2}G(z_2,z_1)^{q/2}\right.\nonumber\\
&\left. -\Sigma(z_1,z_2)G(z_2,z_1) \right]\\
 S_{c}[\Sigma] &= -\ci\ln\int\mathcal{D}(\bar{c},c)\prod_{i\in I}d^{2}\eta_{i}e^{\ci (S_{\Sigma}+S_{in})} \label{Sc}\\
S_{\Sigma}[\bar{c},c,\Sigma] &= \int_{\mathcal{C}} dz_1dz_2\sum_{i}\bar{c}_{i}(z_1)G_{e}^{-1}(z_1,z_2)c_{i}(z_2) \label{eq:SSigma}\\
G_{e}^{-1}&(z_1,z_2) = \mathrm{i} \partial_{z_1}\delta_{\mathcal{C}}(z_1-z_2)-\Sigma(z_1,z_2).\label{SelfCons1}
\end{align}
\end{subequations}
Here $d^2\eta_i=d\bar{\eta}_id\eta_i$ and $S_{in}$ in Eq.\eqref{Sc} above is given in Eq.\eqref{eq:Sin}. 
In the large-$N$ limit, the saddle point equations are obtained by extremizing the action with respect to $G$ and $\Sigma$, leading to
\begin{subequations} \label{eq:Saddle_Fock_A}
\begin{align}
    \Sigma(z_1,z_2) &= J_q^{2}G(z_1,z_2)^{q/2}G(z_2,z_1)^{q/2 - 1} \label{SelfCons2} \\
    G(z_1,z_2) &= -\frac{\mathrm{i}}{N}\sum_{i} \expval{c_{i}(z_1)\bar{c}_{i}(z_2)}_{c}\label{SelfCons3}
\end{align}
\end{subequations}
where the average $\expval{\dots}_{c}$ is evaluated with the fermionic generating function
\begin{align}
Z_c=\int \mathcal{D}(\bar{c},c)\prod_{i\in I}d^2\eta_i e^{\ci(S_\Sigma+S_{in})}, \label{eq:Zc}
\end{align}
for a given $\Sigma(z_1,z_2)$. The latter is self-consistently
determined by $G(z_1,z_2)$ from Eq.\eqref{SelfCons2}. Next,
to evaluate $\langle c_i(z_1)\bar{c}_i(z_2)\rangle_c$ appearing in $G(z_1,z_2)$ [Eq.\eqref{SelfCons3}], and other correlations, we define a generating function $Z_{c}[\bar{\zeta},\zeta]$ with Grassmann source fields $\{\bar{\zeta}_{i}(z),\zeta_{i}(z)\}$
\begin{multline}\label{eq:Zc_zeta}
Z_{c}[\bar{\zeta},\zeta] = \int\mathcal{D}(\bar{c},c)\prod_{i\in I}d^{2}\eta_{i}\exp(\ci (S_{\Sigma}+S_{in}))  \\ \times \exp(\ci \int_{\mathcal{C}} dz\sum_{i}\left[\bar{\zeta}_{i}(z)c_{i}(z)+\bar{c}_{i}(z)\zeta_{i}(z)\right]).
\end{multline}
Using the above, all two-point and higher point correlations can be generated from the expansion in terms of the source fields $\{\bar{\zeta}_{i}(z),\zeta_{i}(z)\}$ as,
\begin{align}
&\frac{Z_c[\bar{\zeta},\zeta]}{Z_c}=1-\sum_{i_1j_1}\int dz_1dz_1'\langle c_{i_1}(z_1)\bar{c}_{j_1}(z_1')\rangle_c \bar{\zeta}_{i_1}(z_1)\zeta_{j_1}(z_1')\nonumber \\
&+\sum_{i_1i_2j_1j_2}\int dz_1dz_2dz_1'dz_2'\langle c_{i_1}(z_1)c_{i_2}(z_2)\bar{c}_{j_2}(z_2')\bar{c}_{j_1}(z_1')\rangle_c\nonumber\\
&~~~~\bar{\zeta}_{i_1}(z_1)\bar{\zeta}_{i_2}(z_2)\zeta_{j_2}(z_2')\zeta_{j_1}(z_1')-\cdots \label{eq:Zc_Exp}
\end{align}
We directly integrate out the fermionic fields $(\bar{c}, c)$ followed by the Grassmann source fields $({\bar{\eta}, \eta})$, encoding the initial-state information, in both $Z_c$ [Eq.\eqref{eq:Zc}] and $Z_c[\bar{\zeta},\zeta]$ [Eq.\eqref{eq:Zc_zeta}] to obtain 
\begin{subequations}
\begin{align}
Z_c&=\mathrm{det}[-\ci G_e^{-1}][\ci G_e(0-,0+)]^{fN} \label{eq:Zc_Expression}\\
Z_c[\bar{\zeta},\zeta]&=Z_c e^{-\ci\int dz_1dz_2\sum_i\bar{\zeta}_i(z_1)G_{ii}(z_1,z_2)\zeta_i(z_2)} \\
G_{ii}(z_1,z_2)&=G_{e}(z_1,z_2) -\delta_{i\in I} \frac{G_{e}(z_1,0+)G_{e}(0-,z_2)}{G_{e}(0-,0+)},\label{SelfCons4}
\end{align}
\end{subequations}
for a given $\Sigma(z_1,z_2)$. Thus comparing $Z_c[\bar{\zeta},\zeta]/Z_c$ above with the expansion in Eq.\eqref{eq:Zc_Exp}, we get $G_{ii}(z_1,z_2)= -\ci\langle c_i(z_1)\bar{c}_i(z_2)\rangle$.
 We see that $G_e(z_1,z_2)$ indeed corresponds to the local Green's function for initially \emph{empty} or unfilled sites in the large-$N$ limit. This allows us to obtain a closed set of large-$N$ self-consistent saddle-point equations. For convenience, we write the full set of self-consistent equations below.
\begin{subequations} \label{eq:SelfConsistency_SYK}
    \begin{align}
  \Sigma(z_1, z_2) &= J_q^{2}G(z_1,z_2)^{q/2}G(z_2,z_1)^{q/2-1} \label{eq:SelfEnergy_SYKq_A}\\
  G(z_1, z_2) &= f G_f(z_1, z_2) + (1-f) G_e(z_1, z_2)\\
    G_{f}(z_1,z_2) &= G_{e}(z_1,z_2) - \frac{G_{e}(z_1,0+)G_{e}(0-,z_2)}{G_{e}(0-,0+)} \label{eq:Gf}\\
    \ci \partial_{z_1}G_e(z_1, z_2) &= \delta_{\mathcal{C}}(z_1 - z_2) + \int_{\mathcal{C}} dz \Sigma(z_1, z) G_e(z, z_2) \\ 
    -\ci \partial_{z_2}G_e(z_1, z_2) &= \delta_{\mathcal{C}}(z_1 - z_2) + \int_{\mathcal{C}} dz G_e(z_1, z) \Sigma(z, z_2) 
    \end{align}
\end{subequations}
where the last two equations are obtained by a convolution of Eq.\eqref{SelfCons1} with $G_e(z_1, z_2)$ from the right and the left respectively. Next, we obtain disorder-averaged non-equilibrium Green's functions in real time. We set $z_1 \rightarrow t_{1} +(t_1-), z_2 \rightarrow t_2 -(t_2+)$ to get the lesser $G^<$ (greater $G^>$) functions. Finally, we use Langreth rules \cite{Stefanucci2013} to convert the integral over contour $\mathcal{C}$ to integral in real time and obtain the following Kadanoff-Baym equations 
\begin{subequations}\label{KBEqns1}
\begin{align}
    \Sigma^{>,<}(t_1, t_2) &= J_q^2 G^{>, <}(t_1, t_2)^{q/2}  G^{<, >}(t_2, t_1)^{q/2-1}\label{KB1}\\
    G_f^{>,<}(t_1, t_2) &= G_e^{>,<}(t_1, t_2) - \frac{G_e^>(t_1, 0)G_e^>(0, t_2)}{G_e^>(0, 0)} \label{KB2}  \\
    G^{>,<}(t_1, t_2) &= f G_f^{>,<}(t_1, t_2) + (1-f) G_e^{>,<}(t_1, t_2) \label{KB3} \\
    \ci \partial_{t_1} G_e^{>, <}(t_1, t_2) &= \int_{0}^{t_1} dt \ \Sigma^R(t_1, t) G_e^{>, <}(t, t_2) \nonumber\\
                    &+ \int_{0}^{t_2} dt \ \Sigma^{>, <}(t_1, t) G_e^A(t, t_2) \label{KB4} \\
    -\ci \partial_{t_2} G_e^{>, <}(t_1, t_2) &= \int_{0}^{t_1} dt \ G_e^R(t_1, t) \Sigma^{>, <}(t, t_2) \nonumber \\
    &+ \int_{0}^{t_2} dt \ G_e^{>, <}(t_1, t) \Sigma^A(t, t_2) \label{KB5},
\end{align}
\end{subequations}
where the retarded and the advanced Green's functions are defined as
\begin{align}
    G^{R}(t_1, t_2) &= \theta(t_{1} - t_{2}) [G^{>}(t_1, t_2) - G^{<}(t_1, t_2)] \\
    G^{A}(t_1, t_2) &= \theta(t_{2} - t_{1}) [G^{<}(t_1, t_2) - G^{>}(t_1, t_2)] 
\end{align}
and similarly for self energy $\Sigma^{R/A}(t_1, t_2)$. The initial conditions at $(t_1, t_2) = (0, 0)$ are 
\begin{subequations}\label{eq:InitialCondition_Ge}
\begin{align}
    G_e^>(0,0) = -\ci &,~~~G_e^<(0,0) = 0,\label{eq:Ge_Initial}\\
    G_f^>(0,0) = 0 &,~~~G_f^<(0,0) = \ci, \\
    G^>(0,0) = \ci (f-1) &,~~~ G^<(0,0) = \ci f. \label{eq:InintialCondition_G}
\end{align}
\end{subequations}

\subsection{General initial pure state} \label{app:ArbPure_SYK}
For a general pure state in a fixed particle number sector, $\ket{\Psi(0)}=\sum_n C_n \ket{n}$, starting with the general SK generating function of Eq.\eqref{eq:SK_Z_ArbPure} and disorder averaging over random SYK couplings, we obtain the same set of equations \eqref{eq:SKAction_SYKFock} and the large-$N$ self-consistency condition of Eqs.\eqref{eq:Saddle_Fock}, except for $S_c$, which is now given by $S_c[\Sigma]=-\ci \ln{Z_c}$, with
\begin{align}
Z_c&=\int \mathcal{D}(\bar{c},c)e^{\ci S_\Sigma}\sum_{nn'}C_nC_{n'}^*\prod_{i\in I_n, i'\in I_{n'}}d\bar{\eta}_{i'}d\eta_i e^{\ci S_{nn'}},
\end{align}
where $S_\Sigma$ is given in Eq.\eqref{eq:SSigma} and $S_{nn'}$ in Eq.\eqref{eq:Snn'}. Similar to Eq.\eqref{eq:Zc_zeta}, we define the generating function with Grassmann source fields $\{\bar{\zeta}_i(z),\zeta_i(z)\}$
\begin{multline}
    Z_{c}[\bar{\zeta},\zeta] = \int\mathcal{D}(\bar{c},c) \sum_{nn'} C_n C^*_{n'} \prod_{\substack{i \in I_n \\ i' \in I_{n'}}} d\bar{\eta}_{i'}d\eta_{i} \\ e^{\ci (S_{\Sigma}+S_{nn'})} e^{\ci \int dz\sum_{i}\left(\bar{\zeta}_{i}(z)c_{i}(z)+\bar{c}_{i}(z)\zeta_{i}(z)\right)}.
\end{multline}
$Z_c$  evaluated to be same as that in Eq.\eqref{eq:Zc_Expression}, while, after somewhat tedious but straightforward Gaussian integrations, we obtain
\begin{subequations}
\begin{align}
&\frac{Z_c[\bar{\zeta},\zeta]}{Z_c}=\sum_n |C_n|^2 e^{-\ci\sum_i\int dz_1dz_2 \bar{\zeta}_i(z_1)G^{(n)}_{ii}(z_1,z_2)\zeta_i(z_2)} \nonumber \\
&+\sum_{n\neq n'}\frac{C_nC_{n'}^*}{[\ci G_e(0-,0+)]^{d_{nn'}fN}}\nonumber\\
&\times  \exp\left[-\ci\sum_{\substack{i\in (I_n\cap I_{n'})\\\cup(\bar{I}_n\cap \bar{I}_{n'})}}\int dz_1dz_2 \bar{\zeta}_i(z_1)G^{(nn')}_{ii}(z_1, z_2)\zeta_i(z_2)\right]\nonumber\\
&\prod_{\substack{i\in I_n\cap \bar{I}_{n'}\\i'\in \bar{I}_n \cap I_{n'}}}\int dz_1dz_2 G_e(z_1,0+)G_e(0-,z_2)\bar{\zeta}_i(z_1)\zeta_{i'}(z_2) \label{eq:ZcbyZ}\\
&G^{(n)}_{ii}(z_1,z_2)=G_e(z_1,z_2)-\delta_{i\in I_n}\frac{G_e(z_1,0+)G_e(0-,z_2)}{G_e(0-,0+)}\\
&G^{(nn')}_{ii}(z_1,z_2)=\delta_{i\in (I_n\cap I_{n'})\cup (\bar{I}_n\cap \bar{I}_{n'})}G_e(z_1,z_2) \nonumber \\
&~~~~~~~~~~~~~~~~~~-\delta_{i\in I_n\cap I_{n'}}\frac{G_e(z_1,0+)G_e(0-,z_2)}{G_e(0-,0+)}
\end{align}
\end{subequations}
In the above, $d_{nn'}N$ is the hamming distance between the Fock states $|n\rangle$ and $|n'\rangle$.
Comparing $Z_c[\bar{\zeta},\zeta]/Z_c$ above with the expansion of Eq.\eqref{eq:Zc_Exp}, we obtain the expressions for $G_{ii}(z_1,z_2)$ and $G_{ij}(z_1,z_2)$ in Eqs.\eqref{eq:LocalG_ArbPure},\eqref{eq:Gij_ArbPure} of the main text.
Any higher-point correlation functions, though cumbersome, can also be obtained in a similar fashion by comparing the expressions for $Z_c[\bar{\zeta},\zeta]/Z_c$ in Eq.\eqref{eq:Zc_Exp} and Eq.\eqref{eq:ZcbyZ}.
From the above, it is easy to show that the large-$N$ Green's function, $G(z_1,z_2)=(1/N)\sum_i G_{ii}(z_1,z_2)=fG_f(z_1,z_2)+(1-f)G_e(z_1,z_2)$, has the same expression as in the case of pure Fock product state, with $G_f(z_1,z_2)$ given in Eq.\eqref{eq:Gf}. Thus, even for an arbitrary initial pure state $|\Psi(0)\rangle=\sum_n C_n |n\rangle$ considered here, the large-$N$ self-consistency equations remain exactly the same as Eqs.\eqref{eq:SelfConsistency_SYK}. Thus, non-equilibrium dynamics of all typical initial pure states in a fixed particle number sector for a particular combination of SYK models is entirely determined by a single universal Green's function, $G_e(z_1,z_2)$. The latter is obtained from a set of universal self-consistency equations \eqref{eq:SelfConsistency_SYK}, along with the initial condition of Eq.\eqref{eq:Ge_Initial}. The relation \eqref{eq:SelfEnergy_SYKq_A} between the self-energy $\Sigma(z_1,z_2)$ and the large-$N$ Green's function $G(z_1,z_2)$ depends on the particular type of SYK model or the combinations of the SYK models.

\redtext{
\section{Proof of instantaneous thermalization of large-$N$ Green's function from Kadanoff-Baym equations}\label{app:G_KB_Thermalization}
\subsection{SYK$_2$}
In this appendix, we prove the  instantaneous thermalization of large-$N$ Green's function, starting from KB equations for SYK$_2$ model. 
We achieve this by showing that the derivative of $G^{>,<}(t_1, t_2)$ with respect to the center of mass time $\mathcal{T} = (t_1 + t_2)/2$ is identically zero.
\begin{align}
    &\ci \partial_{\mathcal{T}}G^{>,<}(t_1, t_2) \\
    &= (\ci \partial_{t_1} + \ci \partial_{t_2})G^{>,<}(t_1, t_2) \\
    &= (\ci \partial_{t_1} + \ci \partial_{t_2})\left(G_e^{>,<}(t_1, t_2) - f \frac{G_e^>(t_1, 0) G_e^>(0, t_2)}{G_e^>(0,0)} \right)\\
    &= \int_{0}^{t_1} dt \ \{ [\Sigma^>(t_1, t) - \Sigma^<(t_1, t)]G_e^{>, <}(t, t_2) \nonumber \\
    & \hspace{2cm} -[G_e^>(t_1, t) - G_e^<(t_1, t) ]\Sigma^{>, <}(t, t_2)\} \nonumber \\
    &+ \int_{0}^{t_2} dt \ \{ \Sigma^{>, <}(t_1, t) [G_e^<(t, t_2) - G_e^>(t, t_2)] \nonumber \\
    & \hspace{2cm}-  G_e^{>, <}(t_1, t) [\Sigma^<(t, t_2) - \Sigma^>(t, t_2)]\} \nonumber \\
    &- \frac{f}{G_e^>(0,0)} \times \nonumber \\
    & \hspace{0.5cm}\left( \int_{0}^{t_1} dt \ [\Sigma^>(t_1, t) - \Sigma^<(t_1, t)]G_e^{>}(t, 0)G_e^>(0, t_2)\right. \nonumber \\
    & \hspace{1cm} \left. - \int_{0}^{t_2} dt \ G_e^>(t_1,0)G_e^{>}(0, t) [\Sigma^<(t, t_2) - \Sigma^>(t, t_2)] \right) \\
    &= \int_{0}^{t_1} dt \ \{ [\Sigma^>(t_1, t) - \Sigma^<(t_1, t)] G^{>, <}(t, t_2) \nonumber\\
    &\hspace{2cm}-[G^>(t_1, t) - G^<(t_1, t) ]\Sigma^{>, <}(t, t_2)\} \nonumber \\
    &+ \int_{0}^{t_2} dt \ \{ \Sigma^{>, <}(t_1, t) [G^<(t, t_2) - G^>(t, t_2)] \nonumber \\
    &\hspace{2cm} -  G^{>, <}(t_1, t) [\Sigma^<(t, t_2) - \Sigma^>(t, t_2)]\}
\end{align}
where in the last line, we used $G^{>,<}(t_1, t_2) = G_e^{>,<}(t_1, t_2) - \frac{f}{G_e^>(0,0)}G_e^>(t_1, 0) G_e^>(0, t_2) $. For $q=2$, $\Sigma^{>,<}(t_1, t_2) = J_2^2 G^{>,<}(t_1, t_2)$ and hence the integrands in the above equation are identically zero. Therefore, the large-$N$ Green's functions $G^{>,<}(t_1, t_2)$ only depend on the relative time $t = t_1 - t_2$. }

\redtext{
\subsection{SYK$_{q \geq 4}$}
In this appendix, we argue the instantaneous thermalization of large-$N$ Green's function, starting from KB equations for SYK$_{q \geq 4}$ model. First, we re-write the Kadanoff-Baym equations entirely in terms of the large-$N$ Green's functions $G^{>,<}(t_1, t_2)$ below, 
\begin{align}
    &\ci \partial_{t_1}G^{>,<}(t_1, t_2)\\
    &= \ci \partial_{t_1}\left(G_e^{>,<}(t_1, t_2) - \frac{f}{G_e^{>,<}(0,0)}G_e^>(t_1,0)G_e^>(0,t_2)  \right) \\
    &= \int_{0}^{t_1} dt \  [\Sigma^>(t_1, t) - \Sigma^<(t_1, t)]G_e^{>, <}(t, t_2) \nonumber \\
    &+ \int_{0}^{t_2} dt \ \{ \Sigma^{>, <}(t_1, t) [G_e^<(t, t_2) - G_e^>(t, t_2)] \nonumber \\
    &- \frac{f}{G_e^>(0,0)}  \int_{0}^{t_1} dt \ [\Sigma^>(t_1, t) - \Sigma^<(t_1, t)]G_e^{>}(t, 0)G_e^>(0, t_2) \\
    &= \int_{0}^{t_1} dt \  [\Sigma^>(t_1, t) - \Sigma^<(t_1, t)]G^{>, <}(t, t_2) \nonumber \\
    &+ \int_{0}^{t_2} dt \  \Sigma^{>, <}(t_1, t) [G^<(t, t_2) - G^>(t, t_2)] \\
    &= \int_{0}^{t_1} dt \  \Sigma^R(t_1, t) G^{>, <}(t, t_2) + \int_{0}^{t_2} dt \  \Sigma^{>, <}(t_1, t) G^A(t, t_2) 
\end{align}
Similarly, it is easy to show that 
\begin{align}
     - \ci \partial_{t_2}G^{>,<}(t_1, t_2) &= \int_{0}^{t_1} dt \ G^R(t_1, t) \Sigma^{>, <}(t, t_2) \nonumber \\
     &+ \int_{0}^{t_2} dt \ G^{>, <}(t_1, t) \Sigma^A(t, t_2)
\end{align}
Therefore, the full set of self-consistency equations is given by 
\begin{align}
    \Sigma^{>,<}(t_1, t_2) &= J_q^2 G^{>,<}(t_1, t_2)^{q/2} G^{<,>}(t_2,t_1)^{q/2 - 1} \\
    \ci \partial_{t_1}G^{>,<}(t_1, t_2) &= \int_{0}^{t_1} dt \  \Sigma^R(t_1, t) G^{>, <}(t, t_2) \nonumber \\
     &+ \int_{0}^{t_2} dt \  \Sigma^{>, <}(t_1, t) G^A(t, t_2) \\
    - \ci \partial_{t_2}G^{>,<}(t_1, t_2) &= \int_{0}^{t_1} dt \ G^R(t_1, t) \Sigma^{>, <}(t, t_2) \nonumber \\
     &+ \int_{0}^{t_2} dt \ G^{>, <}(t_1, t) \Sigma^A(t, t_2)
\end{align}
which are the same set of equations that one would get starting from an infinite temperature thermal state on a KP contour for SYK model. The grand canonical density matrix at infinite temperature is $\rho_\infty = e^{\beta \mu \hat{N}}/Z_\infty$. The initial conditions are $G^>(0,0) = -\ci \expval{c_i c_i^\dagger} = -\ci /[1 + \exp(\beta\mu)], G^<(0,0) = \ci \expval{c_i^\dagger c_i} = \ci /[1 + \exp(-\beta\mu)]$. When the chemical potential is also infinity such that $\exp(\beta\mu) = f/(1-f)$, then the initial conditions of KB equations for $G$ at infinite temperature matches with that of the pure state in Eq.\eqref{eq:InintialCondition_G}. Therefore, the large-$N$ Green's function of a generic pure state instantaneously thermalizes to that of infinite temperature. 
}

\section{Exact diagonalization for non-interacting SYK$_2$ model} \label{ED_SYK2}
In this appendix, we provide the details of the exact diagonalization (ED) calculations for SYK$_2$ model of Eq.\eqref{eq:SYK2}. For a given disorder realization, we diagonalize the $N \times N$ matrix $J_{ij}$ to obtain the eigen energies and wavefunctions, which we denote by $\{\epsilon_\alpha,\psi_\alpha(i)\}$. Since the matrix is $N^2$ in size, we can perform ED for relatively large system sizes compared to the interacting case where the size of the full Hamiltonian matrix is exponential in $N$. We write the Hamiltonian in the diagonal basis as, 
\begin{equation}
    \mathcal{H}_2 = \frac{1}{\sqrt{N}} \sum_{i,j=1}^N J_{ij} c^{\dagger}_i c_j = \sum_{\alpha=1}^N\epsilon_{\alpha}c^{\dagger}_\alpha c_\alpha,
\end{equation}
where $J_{ij}$ are Gaussian random variables with the mean $\expval{J_{ij}} = 0 $, and the variance $\expval{|J_{ij}|^2} = J_2^2$.

We recall that the initial state under consideration is a pure Fock product state $\ket{\Psi(0)} = \ket{n_1 \dots n_N}$ with $n_{i \in I} = 1, n_{i \in \bar{I}} = 0, N_I = fN$ where $f$ is the average density or the filling of the initial state. To study the time evolution of initially filled ($i\in I$) and unfilled ($i\in \bar{I}$) sites, we express the on-site density as a function of time in the diagonal basis, 
\begin{equation}\label{cdagtct}
\expval{c^\dagger_i(t) c_i(t)} = \sum_{\alpha \beta} \psi^*_\beta(i) \psi_\alpha(i)    \expval{e^{\ci \mathcal{H}_2 t} c^\dagger_\beta c_{\alpha}e^{-\ci \mathcal{H}_2 t}},
\end{equation}
where $\expval{\bullet}=\bra{\Psi(0)}\bullet\ket{\Psi(0)}$. Using the Heisenberg's equation of motion, we get $c_{\alpha}(t) = e^{\ci \mathcal{H}_2 t} c_\alpha e^{-\ci \mathcal{H}_2 t} =e^{-\ci \epsilon_\alpha t}c_\alpha $.
Expressing the operators in diagonal basis back to the site basis and using $\expval{c^\dagger_i c_j} = n_i \delta_{ij}$, we obtain
\begin{equation}
    \expval{c^\dagger_i(t) c_i(t)} = \sum_j \sum_{\alpha \beta}  \psi_\alpha(i) \psi_\beta^*(i) e^{-\ci(\epsilon_\alpha - \epsilon_\beta)t} \psi^*_\alpha(j) \psi_\beta(j) n_j. 
\end{equation}
We define the densities of initially filled and unfilled sites respectively as, 
\begin{align}
    n_f(t) &= \frac{1}{fN}\sum_{i \in I} \overline{\expval{c^\dagger_i(t) c_i(t)}} \\
    n_e(t) &= \frac{1}{(1-f)N}\sum_{i \notin I} \overline{\expval{c^\dagger_i(t) c_i(t)}} \label{nc}
\end{align}
where $\overline{(\dots)}$ is average over many disorder realizations of the random hopping $J_{ij}$. The results obtained from the above are plotted in Fig.\ref{fig:SYKHalfFilling}(a) as a benchmark of the large-$N$ results for $n_{e,f}(t)$.

\section{Large-$q$ SYK} \label{App:largeqSYK}
In this appendix, we provide the equations and steps supporting the derivation of large-$q$ solution in the main text. The $1/q$ expansions of the Green's functions are given by
\begin{subequations}\label{largeqG}
\begin{align}
    G^<(t_1, t_2) &= \ci f \left(1 + \frac{g^<(t_1, t_2)}{q} +  O(q^{-2}) \right) \\
    G^>(t_1, t_2) &= \ci (f-1) \left(1 + \frac{g^>(t_1, t_2)}{q} + O(q^{-2})\right)  \\
    G_e^<(t_1, t_2) &= \ci \left( \frac{g_e^<(t_1, t_2)}{q} + O(q^{-2})\right) \\
    G_e^>(t_1, t_2) &= -\ci  \left(1 + \frac{g_e^>(t_1, t_2)}{q} + O(q^{-2})\right)  \\
    G_f^<(t_1, t_2) &= \ci \left(1+ \frac{g_f^<(t_1, t_2)}{q} + O(q^{-2})\right) \\
    G_f^>(t_1, t_2) &= \ci  \left( \frac{g_f^>(t_1, t_2)}{q} + O(q^{-2})\right)
\end{align}
\end{subequations}
The Green's functions are not all independent but are related through Eqs.\eqref{KB2} and \eqref{KB3}. Requiring these relations to hold at order $1/q$ yields
\begin{subequations}\label{eq:largeq_g}
\begin{align} 
    g_f^>(t_1, t_2) &= -g_e^>(t_1, t_2) + [g_e^>(t_1,0) + g_e^>(0,t_2)]\\
    g_f^<(t_1, t_2) &= g_e^<(t_1, t_2) + [g_e^>(t_1,0) + g_e^>(0,t_2)]\\
    g^>(t_1, t_2) &=-\frac{f}{1-f} g_f^>(t_1,t_2) + g_e^>(t_1, t_2) \\
    g^<(t_1, t_2) &= g_f^<(t_1,t_2) +\frac{1-f}{f}  g_e^<(t_1, t_2)
\end{align}
\end{subequations}
The KB equations \eqref{KB4} and \eqref{KB5} at order $1/q$ are given by 
\begin{subequations}\label{eq:largeq_KBg}
\begin{align}
    \partial_{t_1}g_e^>(t_1, t_2) & = 2\mathcal{J}^2 (1-f) \int_{t_1}^{t_2}dt e^{g_+(t_1,t)} \nonumber\\
    &- 2\mathcal{J}^2 f \int_{0}^{t_1}dt e^{g_+(t,t_1)}\label{gcg1}\\
   \partial_{t_2}g_e^>(t_1, t_2) & = -2\mathcal{J}^2 (1-f) \int_{t_1}^{t_2}dt e^{g_+(t,t_2)} \nonumber \\ &- 2\mathcal{J}^2 f \int_{0}^{t_2}dt e^{g_+(t_2,t)} \\
    \partial_{t_1}g_e^<(t_1, t_2) & = 2\mathcal{J}^2 f \int_0^{t_2} dt e^{g_+(t,t_1)}\label{gcl1} \\
    \partial_{t_2}g_e^<(t_1, t_2) & =  2\mathcal{J}^2 f \int_0^{t_1} dt e^{g_+(t_2,t)}.
\end{align}
\end{subequations}
Differentiating Eqs.\eqref{gcg1}, \eqref{gcl1} with $t_2$ gives 
\begin{align}
    \partial_{t_2}\partial_{t_1}g_e^>(t_1, t_2) &= 2\mathcal{J}^2 (1-f) e^{g_+(t_1,t_2)} \\
    \partial_{t_2}\partial_{t_1}g_e^<(t_2, t_1) &= 2\mathcal{J}^2 f e^{g_+(t_1,t_2)}.
\end{align}
Using the above two equations along with the definition of $g_+(t_1, t_2)= (g^>(t_1, t_2) + g^<(t_2, t_1))/2 = g_e^>(t_1, t_2)/(2-2f) + g_e^<(t_2, t_1)/(2f)  -f[g_e^>(t_1, 0) + g_e^>(0, t_2)]/(2-2f) +  [g_e^>(t_2, 0) + g_e^>(0, t_1)]/2 $ gives us the Liouville equation \eqref{LiouvilleEquation} in the main text. 
\begin{figure}[H]
    \centering
    \includegraphics[width=\linewidth]{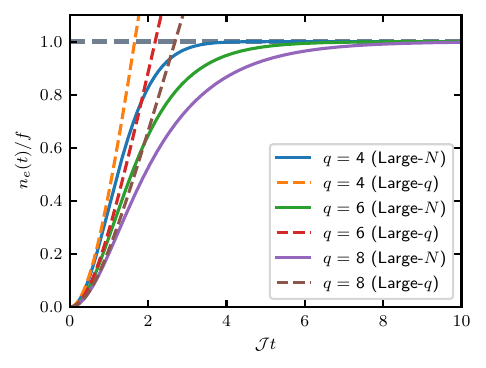}
    \caption{Comparison of large-$N$ Kadanoff-Baym predictor-corrector numerics with the large-$q$ analytical solution, $n_e(t) =f (4/q) \cosh{\mathcal{J} t}$, where $\mathcal{J}  = J_q \sqrt{(q/2)} (f-f^2)^{q/4 -1/2}$, for various values of $q$: higher values of $q$ show better agreement with numerics for a longer time.}
    \label{fig:SYKqScaling}
\end{figure}





\bibliography{SYKPureState}

\end{document}